\documentclass[journal]{IEEEtran}
\usepackage[table]{xcolor}
\usepackage[nocompress]{cite}
\usepackage{amsmath,amssymb,amsfonts}
\usepackage{algorithmic}
\usepackage{graphicx}
\usepackage{textcomp}
\usepackage{bm}
\usepackage{soul}
\usepackage{xcolor} 

\makeatletter
\AtBeginDocument{\DeclareMathVersion{bold}
\SetSymbolFont{operators}{bold}{T1}{times}{b}{n}
\SetMathAlphabet{\mathrm}{bold}{T1}{times}{b}{n}
\SetMathAlphabet{\mathit}{bold}{T1}{times}{b}{it}
\SetMathAlphabet{\mathbf}{bold}{T1}{times}{b}{n}
\SetMathAlphabet{\mathtt}{bold}{OT1}{pcr}{b}{n}
\SetSymbolFont{symbols}{bold}{OMS}{cmsy}{b}{n}
\renewcommand\boldmath{\@nomath\boldmath\mathversion{bold}}}
\makeatother

\def\BibTeX{{\rm B\kern-.05em{\sc i\kern-.025em b}\kern-.08em
    T\kern-.1667em\lower.7ex\hbox{E}\kern-.125emX}}

\begin{document}
\title{Study of Switched Step-size Based Filtered-x NLMS Algorithm for Active Noise Cancellation}
\author{Zhiyuan Li, Yi Yu, Hongsen He, Yuyu Zhu and Rodrigo C. de Lamare
\thanks{The first authors are with the School of Information and Control Engineering, Southwest University of Science and Technology, Mianyang, 621010, China (e-mail: li\_zhi\_yuan\_1@163.com, yuyi\_xyuan@163.com, hongsenhe@gmail.com, zhuyuyu@swust.edu.cn). R. de Lamare is with CETUC, PUC-Rio, Rio de Janeiro 22451-900, Brazil (e-mail: delamare@puc-rio.br)}}

\maketitle
  
%\tfootnote{This work was supported in part by the National Natural Science Foundation of China (Nos. 62471412 and 62071399) and supported in part by the Central Guidance for Local Scientific and Technological Development Foundation (2025ZYDF004).}

%\markboth
%{Author \headeretal: Preparation of Papers for IEEE TRANSACTIONS and JOURNALS}
%{Author \headeretal: Preparation of Papers for IEEE TRANSACTIONS and JOURNALS}

%\corresp{Corresponding author: Yi Yu (e-mail: yuyi\_xyuan@163.com), Yuyu Zhu (e-mail: zhuyuyu@swust.edu.cn).}

\begin{abstract}
While the filtered-x normalized least mean square (FxNLMS) algorithm is widely applied due to its simple structure and easy implementation for active noise control system, it faces two critical limitations: the fixed step-size causes a trade-off between convergence rate and steady-state residual error, and its performance deteriorates significantly in impulsive noise environments. To address the step-size constraint issue, we propose the switched \mbox{step-size} FxNLMS (SSS-FxNLMS) algorithm. Specifically, we derive the \mbox{mean-square} deviation (MSD) trend of the FxNLMS algorithm, and then by comparing the MSD trends corresponding to different \mbox{step-sizes}, the optimal step-size for each iteration is selected. Furthermore, to enhance the algorithm's robustness in impulsive noise scenarios, we integrate a robust strategy into the SSS-FxNLMS algorithm, resulting in a robust variant of it. The effectiveness and superiority of the proposed algorithms has been confirmed through computer simulations in different noise scenarios.	
\end{abstract}

%\titlepgskip=-21pt

\maketitle

\section{Introduction}
\label{sec:introduction}
The rapid progress of urbanization has not only brought convenience to people's lives but also exposed them to severe more noise pollution, such as industrial noise, traffic noise and society noise. When staying in noisy surroundings for a long period, people's work efficiency will be significantly decreased and their physical and mental health will suffer from serious negative impacts. To provide a quiet auditory environment for humans, noise cancellation technologies have come into existence. Currently, they can mainly divide into two categories: passive noise control (PNC) and active noise control (ANC). In the implementation of PNC technology, sound-absorbing or sound-insulating materials are generally needed, with its structure being rather bulky. It workes well in eliminating medium and high frequency noises while having poor performance decreasing low frequency noises. On the contrary, ANC shows outstanding efficacy for controling the low frequency noise and has become a research hotspot in related fields nowadays\cite{shi2021optimal, 10806852}. These significant advancements in signal processing algorithms and applications over recent decades have been comprehensively reviewed in \cite{shi2023active}. The ANC technology extensively applied in scenarios such as vehicle interior\cite{zhang2019normalized}, headsets\cite{zhang2014causality}, home appliances\cite{lam2020active, mazur2019active, lee2021review}, construction machines\cite{wen2021design, shi2022integration} and pipelines\cite{wang2025vibration}.

The realization of ANC relies on the principle of sound wave interference cancellation\cite{aboutiman2025active}. A reference microphone sensor is arranged near the noise source, which picks up the noise signal and then transmits it to the noise controller. After being processed by the noise controller, the signal is played through the secondary loudspeaker, resulting in the generation of the cancellation signal with an opposite phase. The original sound and the cancellation sound interfere with each other at \mbox{noise-cancellation} point to yield a quiet area. At the same time, the residual noise is detected in \mbox{real-time} by the microphone error sensor located in the noise-reduction area and fed back to the noise controller to adjust the adaptive filter's coefficients in the controller, so as to achieve more efficient and accurate noise reduction\cite{hamada1991signal,kajikawa2012recent}. 

The performance of ANC depends mainly on the adaptive learning algorithms \cite{aifir,jio,jidf} in the controller. The filtered-x least mean square (FxLMS) algorithm is widely used for ANC systems\cite{kuo1995adaptive,morgan2013history,zhu2023fractional}. However, there exists a problem that it fails to give consideration to both convergence rate and steady-state residual error simultaneously. In order to address this problem, many step-size adjustment strategies have been proposed\cite{song2019filtered,zhou2024combined,kar2024improved,gomathi2016variable}. In \cite{song2019filtered}, the convex combination of the filtered-x least mean square/fourth (CC-FxLMS/F) algorithm was developed, which integrated the performance merits of two filters with different step-sizes through the mixing parameter. The CC family has difficulty in choosing suitable mixing parameter, and it requires an additional filter, which increases the computational complexity. The idea of combined step-size (CSS) is also to combine characteristics of large \mbox{step-size} and small step-size\cite{zhou2024combined}, but differing from \mbox{CC-type}, the CSS algorithm updates the single filter by adjusting the ratio of two step-sizes through the mixing parameter. As another alternative scheme, the variable step-size (VSS) replaces the fixed step-size with an adjustable step-size at each iteration.  In \cite{kar2024improved}, the adaptation step-size was adjusted dynamically according to the input signal and the instantaneous error of the controller. In \cite{gomathi2016variable}, the authors provided a VSS algorithm based on arctangent function of the error signal. However, in complex noise environments, the instability of error signals may lead to incorrect step-size selection adjustment in the aforementioned VSS algorithms.

It is notable that the impulsive noise exists in practical environments, such as the sound of train horns, pile drivers, and firecrackers\cite{wang2023recursive}. As there is no bounded second moment for impulsive noises, the traditional FxLMS algorithm based on the minimum mean square error criterion is at risk of the deterioration of noise reduction performance and instability\cite{pawelczyk2015extension}. In the literature, many ANC algorithms have been proposed to cope with the effect of impulsive noises, and these algorithms can be classified into two types. The one is to clip samples of the reference signal when their amplitudes go beyond a specific threshold, such as the M-estimator based FxLMS algorithm\cite{wu2013m} and the soft-threshold based algorithm regarding reference and error signals\cite{saravanan2019active}. The other is that different robust continuous cost functions were utilized to derive new ANC algorithms, such as the maximum versoria criterion (MVC)\cite{maya2023new,cheng2022active}, minimum error p-power criterion\cite{yang2023robust}, exponential hyperbolic cosine (EHCF) norm\cite{kumar2024robust} and maximum correntropy criterion (MCC)\cite{zhu2020robust,kurian2017robust}. 
To further enhance the performance of ANC algorithms based on robust cost functions, extensive studies have been carried out.
In \cite{chien2022affine}, it combined the MCC with the affine projection framework by exploiting multiple consecutive error samples.
A hyperbolic tangent exponential kernel was incorporated into the M-estimator objective function in \cite{hermont2025robust}.
A generalized robust loss function proposed in \cite{barron2019general} was employed in \cite{zhou2025active} to construct the cost function. This loss function was a superset of several classical robust loss functions and was minimized to achieve robust optimization.
In addition, there exist other robust approaches in adaptive filtering that can provide useful insights for the ANC field, such as mixed p-norm–based methods\cite{zayyani2014continuous,zayyani2024robust,habibi2021robust}, maximum a posteriori estimation approaches \cite{habibi2023robust}, and robust diffusion adaptive filtering algorithms \cite{9050865}, \cite{8618326}. However, these robust algorithms for combating impulsive noises still cannot avoid the tradeoff problem caused by the fixed step-size. In \cite{maya2023new}, it exploited the advantages of two MVC filters with distinct parameters, and dynamically switched between them for updates according to a selection criterion. Nevertheless, choosing the mixing parameter for this approach is challenging.

In this paper, the method of switched step-size (SSS) is employed to address the tradeoff problem caused by the fixed step-size in the ANC algorithm. This is an innovative method proposed by our group in 2021, whose principle is to establish a mean-square deviation (MSD) recursion to describe the evolution behavior of the algorithm with iterations, and then by comparing MSD trends corresponding to different \mbox{step-sizes}, the step-size associated with the smallest MSD will be the optimal step-size for the current iteration. Essentially, it is different from the CC, CSS and VSS technologies. Thus, the SSS has been applied to enhance the performance of algorithms in system identification\cite{guo2022normalized,huang2022general}, echo cancellation\cite{li2024frequency} and beamforming\cite{li2024switching}, yet it has not been explored in the ANC field. Moreover, a switching-based variable step-size approach was proposed for the conventional LMS algorithm in \cite{chien2013switching}, where the step-size is computed by minimizing the MSD and adjusted at different switching points based on noise-free error estimation. By contrast, our proposed method is MSD-driven and directly switches the step-size by selecting the optimal value from a predefined set, and is developed for the FxNLMS algorithm in ANC systems.

The contributions of this paper are as follows. First, we derive the MSD recursion that models the evolution trend of the filtered-x normalized LMS (FxNLMS)  algorithm,  which reveals the step-size's influence law on the algorithm performance,  and then propose the SSS-based FxNLMS algorithm based on this recursion to enhance the performance of the ANC controller in terms of convergence rate and steady-state residual error. Second, to solve the problem of the impulsive noise, we incorporate the robust strategy into the \mbox{SSS-FxNLMS} algorithm, which yields the robust SSS-FxNLMS (\mbox{R-SSS-FxNLMS}) algorithm.

Although reference \cite{guo2020convergence} proposed a relative accuracy recursive MSD model for the FxNLMS algorithm, it depended on the Kronecker product operation which required high computational complexity with the amount of $\mathcal{O}(L^2 \times L^2)$, where $L$ represents the length of the adaptive filter. Thus, the existing MSD recursion is not suitable for choosing the step-size in the SSS scheme. For that reason, this paper establishes the weight deviation covariance matrix, simplifies and analyzes it using some approximations and assumptions. Then, vectorizing and summing the matrix to provide a directly computable formula for the MSD recursion. 

The structure of this paper is arranged as follows. \mbox{Section \mbox{\uppercase\expandafter{\romannumeral2}}} reviews the FxNLMS algorithm. In Section \mbox{\uppercase\expandafter{\romannumeral3}}, we establish the recursion MSD model of FxNLMS, and propose the \mbox{SSS-FxNLMS} and \mbox{R-SSS-FxNLMS} algorithms. Section \mbox{\uppercase\expandafter{\romannumeral4}} presents simulation results to evaluate the algorithms' performance in different noise scenarios. Finally, the conclusion is given. 

\section{Review of FxNLMS}
\begin{figure}[htbp]
	\centering
	{
		\begin{minipage}[a]{0.45\textwidth}
			\includegraphics[width=1\textwidth]{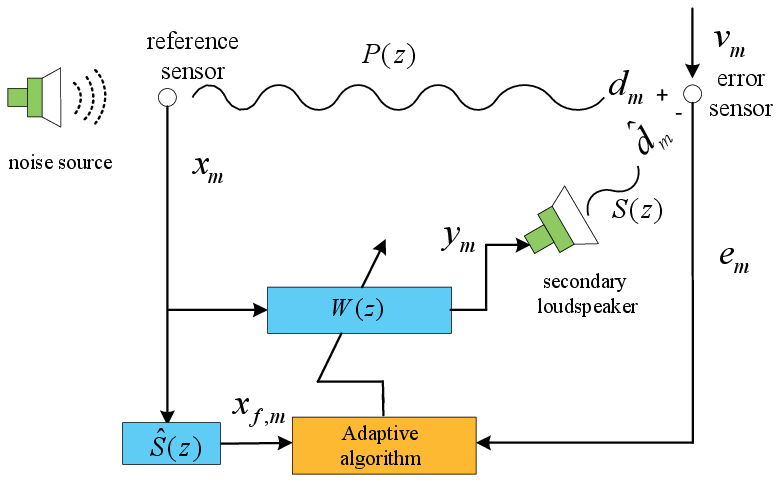} 
		\end{minipage}
	}
	\caption{Diagram of a single-channel feedforward ANC system.}
	\label{Fig1}
\end{figure}
The structural diagram of the single channel feedforward ANC system is shown in Fig.1, where $P(z)$, $S(z)$ and $W(z)$ denote the transfer functions of the primary path, the secondary path and the adaptive filter respectively. The noise source emits the primary noise, and it is picked up by the reference microphone sensor to obtain the reference input signal $x_{m}$. The noise source signal $x_{m}$ passes through $P(z)$ to generate the disturbance noise signal $d_{m}$, which has the following linear relationship
\begin{equation}
	d_{m}=x_{m}*{p}_{m},
	\label{eq001}
\end{equation}
where $p_{m}$ denotes the impulse response of the primary path and $*$ represents the convolution operation. The purpose of ANC technology is to eliminate the disturbance noise signal $d_{m}$. To achieve this goal, the reference input signal $x_{m}$ goes through the adaptive filter $W(z)$ to obtain an secondary sound source signal, described as
\begin{equation}
	y_{m}=\textbf{x}^{\text{T}}_{m}\textbf{w}_{m},
	\label{eq002}
\end{equation}
where $\textbf{x}_{m}=[x_{m},x_{m-1},\cdots,x_{m-L+1}]^{\text{T}}$ is the reference input vector, $(\cdot)^{\text{T}}$ means the mathematical transpose and  $\textbf{w}_{m}$ represents the weight vector of $W(z)$ with the length of $L$. 
Theoretically, we hope that the output signal $y_{m}$ can cancel out the disturbance signal $d_{m}$. However, it is impossible because of the existence of the secondary path $S(z)$ between the secondary sound source and the error microphone sensor in practice. It includes the loudspeaker, analog-to-digital conversion circuit, digital-to-analog conversion circuit, reconstruction filter, error microphone sensor, anti alias filter, power amplifier, preamplifier, and acoustic path from the loudspeaker to the error microphone sensor\cite{Kuo1996ActiveNC}. Thus, after $y_{m}$ is played through the loudspeaker, it will drives a secondary cancellation signal
\begin{equation}
	\hat{d}_{m}=\textbf{y}^{\text{T}}_{m}\hat{\textbf{s}},
	\label{eq003}
\end{equation}
where $\textbf{y}_{m}=[y_{m},y_{m-1},\cdots,y_{m-M+1}]$ and $\hat{\textbf{s}}=[\hat{s}_{0},\hat{s}_{1},\cdots,\hat{s}_{M-1}]$ denotes the impulse response of the secondary path with the length of $M$. The secondary path is unknown, but generally it can be estimated using online or offline modeling methods\cite{chan2011performance,8472270,7760885}. Here, we assume that $S(z)$ has been simulated completely by $\hat{S}(z)$. Subsequently, $d_{m}$ and $\hat{d}_{m}$ interact with each other to yield a quiet area at the noise-cancellation point, and the residual error $e_{m}$ after the interaction is gathered by the error microphone sensor,  given by
\begin{equation}
	e_{m}=d_{m}-\hat{d}_{m}+v_{m},
	\label{eq004}
\end{equation}
where $v_{m}$ represents the measurement noise.

The ANC algorithm keeps adjusting the weights of $W(z)$ to make $\hat{d}_{m}$ as close as possible to ${d}_{m}$, thereby continuously and effectively reducing $e_{m}$. The weight update of the FxNLMS algorithm is expressed as\cite{10443374}
\begin{equation}
	\textbf{w}_{m+1}=\textbf{w}_{m}+\frac{\mu\textbf{x}_{f,m}e_{m}}{||\textbf{x}_{f,m}||_{2}^{2}},
	\label{eq005}
\end{equation}
where $\mu$ is the step-size and $\textbf{x}_{f,m}=[{x}_{f,m},{x}_{f,m-1},\cdots,{x}_{f,m-L+1}]$ is the filtered reference vector whose elements are obtained by 
\begin{equation}
	{x}_{f,m}=\sum_{i = 0}^{M-1}\hat{s}_{i}x(m - i).
	\label{eq006}
\end{equation}

\section{Proposed Algorithms}
In this section, we first present the performance analysis of the FxNLMS algorithm, and then provide deployments of the \mbox{SSS-FxNLMS} and R-SSS-FxNLMS algorithms based on the analytical results.
 \subsection{Performance analysis of FxNLMS}
 \label{sec:Performance analysis of FxNLMS}
As mentioned before, there is a secondary path ${S}(z)$ after the adaptive filter ${W}(z)$, which increases the difficulty of establishing MSD of the algorithm. If the order of ${S}(z)$ and ${W}(z)$ are exchanged, the analysis will be simplified. Fortunately, it is feasible because of the commutativity of convolution and the assumption of slow convergence. Moreover, this treatment has been frequently utilized in the design and analysis of ANC algorithms \cite{resende2005efficient,resende2006statistical,guo2020convergence}. The slow convergence assumption indicates that ${W}(z)$ keeps a constant during the past $M$ sampling periods, meaning $\textbf{w}_{m}\approx\textbf{w}_{m-1}\approx\cdots\approx\textbf{w}_{m-M+1}$\cite{yang2020stochastic}. Based on the above consideration, one can assume that  there is an optimal weight $\textbf{w}_{o}$ of the adaptive filter, and thus the desired signal given in $\left(\ref{eq001}\right)$ can be rearranged as
\begin{equation}
	d_{m}=\textbf{x}_{f,m}^{\text{T}}\textbf{w}_{o}.
	\label{eq007}
\end{equation}
Accordingly, the error signal $e_{m}$ can be changed as
\begin{equation}
	\begin{aligned}
		e_{m}&=\textbf{x}_{f,m}^{\text{T}}\textbf{w}_{o}-\textbf{x}_{f,m}^{\text{T}}\textbf{w}_{m}+v_{m}\\
		&=\textbf{x}_{f,m}^{\text{T}}\widetilde{\textbf{w}}_{m}+v_{m},
		\label{eq008}
	\end{aligned}
\end{equation}
where $\widetilde{\textbf{w}}_{m}\triangleq\textbf{w}_{o}-\textbf{w}_{m}$ defines the weight deviation vector.

Subtracting both side of $\left(\ref{eq005}\right)$ by $\textbf{w}_{o}$, we obtain
\begin{equation}
	\widetilde{\textbf{w}}_{m+1}=\widetilde{\textbf{w}}_{m}-\mu\frac{\textbf{x}_{f,m}e_{m}}{||\textbf{x}_{f,m}||_{2}^{2}}.
	\label{eq009}
\end{equation}
1) Mean analysis\\

To analyze the convergence behavior of the FxNLMS algorithm in the mean sense, we substitute \eqref{eq008} into \eqref{eq009} and take the expectation on both sides, which yields

\begin{equation}
	\text{E}\left\{\widetilde{\textbf{w}}_{m+1}\right\}=\text{E}\left\{\widetilde{\textbf{w}}_{m}-\mu\frac{\textbf{x}_{f,m}(\textbf{x}_{f,m}^{\text{T}}\widetilde{\textbf{w}}_{m}+v_{m})}{||\textbf{x}_{f,m}||_{2}^{2}}\right\}.
	\label{eq010}
\end{equation}

To handle the mathematical expectations involved in $\left(\ref{eq010}\right)$, we need the following assumptions. 

Assumption 1: The measurement noise $v_{m}$ is subject to a Gaussian distribution with zero-mean and variance $\sigma_{v}^{2}$, which is dependent of other signals.

Assumption 2: $\textbf{x}_{f,m}$ and $\widetilde{\textbf{w}}_{m}$ are independent of each other. This assumption has been widely adopted in both design and analysis of adaptive filters to yield satisfactory results\cite{sayed2003fundamentals,farhang2013adaptive}. Also, its validity in ANC has been examined in some references \cite{guo2020convergence, yang2020stochastic}.

Under the above assumptions, $\left(\ref{eq010}\right)$ can be written as
\begin{equation}
	\text{E}\left\{\widetilde{\textbf{w}}_{m+1}\right\}=\left(\textbf{I}-\mu\bm{\Lambda}\right)\text{E}\left\{\widetilde{\textbf{w}}_{m}\right\},
	\label{eq011}
\end{equation}
where $\bm{\Lambda}=\text{E}\left\{\frac{\textbf{x}_{f,m}\textbf{x}_{f,m}^{\text{T}}}{||\textbf{x}_{f,m}||_{2}^{2}}\right\}$, and $\text{E}\left\{\cdot\right\}$ denotes the mathematical expectation.

For mean convergence, the spectral radius of matrix $\left(\textbf{I}-\mu\frac{\textbf{x}_{f,m}\textbf{x}_{f,m}^{\text{T}}}{||\textbf{x}_{f,m}||_{2}^{2}}\right)$ must be smaller than $1$, i.e., $\left| \textbf{I}-\mu\frac{\textbf{x}_{f,m}\textbf{x}_{f,m}^{\text{T}}}{||\textbf{x}_{f,m}||_{2}^{2}}\right|<1$\cite{farhang2013adaptive}. Then, the step-size must satisfy
\begin{equation}
0<\mu<\frac{2}{\lambda_{max}(\bm{\Lambda})},
	\label{eq012}
\end{equation}
where $\lambda_{\max}(\cdot)$ represents to take the largest eigenvalue.

2) Mean-square analysis

By post-multiplying both sides of $\left(\ref{eq009}\right)$ by its transpose and then taking expectations of all the terms, we can get
\begin{equation}
	\begin{aligned}
		\textbf{P}_{m+1}=&\textbf{P}_{m}-\mu\text{E}\left\{\frac{\textbf{x}_{f,m}e_{m}}{||\textbf{x}_{f,m}||_{2}^{2}}\widetilde{\textbf{w}}^{\text{T}}_{m}\right\}-\mu\text{E}\left\{\widetilde{\textbf{w}}_{m}\frac{\textbf{x}_{f,m}^{\text{T}}e_{m}}{||\textbf{x}_{f,m}||_{2}^{2}}\right\}\\
		&+\mu^{2}\text{E}\left\{e^{2}_{m}\frac{\textbf{x}_{f,m}\textbf{x}_{f,m}^{\text{T}}}{||\textbf{x}_{f,m}||_{2}^{4}}\right\},
		\label{eq013}
	\end{aligned}
\end{equation}
where $\textbf{P}_{m}\triangleq\text{E}\left\{\widetilde{\textbf{w}}_{m}\widetilde{\textbf{w}}^{\text{T}}_{m}\right\}$ denotes the covariance matrix of $\widetilde{\textbf{w}}_{m}$.

Based on some approximations and assumptions, $\left(\ref{eq013}\right)$ can be simplified as (Seeing Appendix A for its detial)

\begin{equation}
	\begin{aligned}
		\textbf{P}_{m+1}=&\textbf{P}_{m}-2\mu\frac{\textbf{R}_{f,m}\textbf{P}_{m}}{||\textbf{x}_{f,m}||_{2}^{2}}+\mu^{2}\frac{\sigma_{e}^{2}\textbf{R}_{f,m}}{||\textbf{x}_{f,m}||_{2}^{4}}\\
		&+2\mu^{2}\frac{\textbf{R}_{f,m}\textbf{P}_{m}\textbf{R}_{f,m}}{||\textbf{x}_{f,m}||_{2}^{4}}+\mu^{2}\frac{\textbf{R}_{f,m}\text{Tr}[\textbf{R}_{f,m}\textbf{P}_{m}]}{||\textbf{x}_{f,m}||_{2}^{4}},
		\label{eq014}
	\end{aligned}
\end{equation}
where $\textbf{R}_{f,m}\triangleq\textbf{x}_{f,m}\textbf{x}_{f,m}^{\text{T}}$.

According to the definition of MSD, denoted as
\begin{equation}
	\begin{aligned}
		J_{m}&\triangleq\text{E}\left\{||\textbf{w}_{m}-\textbf{w}_{o}||^{2}_{2}\right\}\\
		&=\text{Tr}[\textbf{P}_{m}],
		\label{eq015}
	\end{aligned}
\end{equation}
we can decribe the evolution behavior of the FxNLMS algorithm with iterations by using \eqref{eq014}.

It is noticed that there are numerous matrix operations involved in $\left(\ref{eq014}\right)$, which could have two challenges. For one thing, the matrix-based update exists numerical instability, due to the potential ill-condition coupled with cumulative rounding errors in finite-precision arithmetic, which aligns with the longstanding issue observed in the recursive least squares\cite{arablouei2014unbiased, zakharov2008low}. For another thing,  the matrix update results in a high computational complexity. Given that our improvement focuses on employing the evolution trend of MSD rather than its exact value, the vectorized approximation of $\textbf{P}_{m}$ is enough for step-size selection while ensuring numerical robustness, so that we can neglect the \mbox{non-diagonal} elements of $\textbf{R}_{f,m}$ and $\textbf{P}_{m}$. By doing so, a low complexity version of \eqref{eq014} is obtained as
\begin{equation}
	\begin{aligned}
		\overline{\textbf{P}}_{m+1}=&(1-2\mu\frac{\overline{\textbf{R}}_{f,m}}{||\textbf{x}_{f,m}||_{2}^{2}})\odot\overline{\textbf{P}}_{m}+\mu^{2}\frac{\overline{\textbf{R}}_{f,m}\text{sum}[\overline{\textbf{R}}_{f,m}\odot\overline{\textbf{P}}_{m}]}{||\textbf{x}_{f,m}||_{2}^{4}}\\
		&+2\mu^{2}\frac{\overline{\textbf{R}}_{f,m}\odot\overline{\textbf{P}}_{m}\odot\overline{\textbf{R}}_{f,m}}{||\textbf{x}_{f,m}||_{2}^{4}}+\mu^{2}\frac{\sigma_{e}^{2}\overline{\textbf{R}}_{f,m}}{||\textbf{x}_{f,m}||_{2}^{4}}
,
		\label{eq016}
	\end{aligned}
\end{equation}
where $\overline{\textbf{P}}_{m}$ and $\overline{\textbf{R}}_{f,m}$ are vectors formed by the diagonal elements of ${\textbf{P}}_{m}$ and $\textbf{R}_{f,m}$, respectively,  $\text{sum}[\cdot]$ denotes the sum of all elements in a vector, and  $\odot$ represents the element-product of two vectors. Accordingly, the MSD trend of the FxNLMS algorithm can be calculated by
\begin{equation}
	J_{m+1}\triangleq\text{sum}[\overline{\textbf{P}}_{m+1}].
	\label{eq017}
\end{equation}

\vspace{0.4cm}

\textbf{Remark 1:} To discuss the convergence of the  algorithm with respect to the step-size, we further assume that the filtered input signal $x_{f,m}$ is white process with a finite second-order moment \cite{farhang2013adaptive, sayed2003fundamentals}. Thus,we have $\overline{\textbf{R}}_{f,m}=\sigma_{f}^{2}\textbf{I}_{L\times1}$ and $||\textbf{x}_{f,m}||_{2}^{2}=L\sigma_{f}^{2}$, where $\sigma_{f}^{2}$ represents the variance of $x_{f,m}$ and $\textbf{I}_{L\times1}$ is a column vector of length $L$ with all elements being 1. From a strict perspective, this assumption may not necessarily meet the actual situation. However, considering that our goal is only to explain the algorithm's convergence controlled by the fixed step-size, this assumption is valid and has been widely adopted in the design and analysis of adaptive algorithms\cite{2016An,sulyman2003convergence}. Based on this assumption, we reconstruct \eqref{eq017} as
\begin{equation}
	\begin{aligned}
		J_{m+1}=h_{m}J_{m}+\mu^{2}\frac{\sigma_{e}^{2}}{L^{2}\sigma_{f}^{2}},
		\label{eq018}
	\end{aligned}
\end{equation}
where 
\begin{equation}
	h_{m}=1-2\mu\frac{1}{L}+\mu^{2}\frac{L+2}{L^{2}}.
	\label{eq019}
\end{equation}

Equation \eqref{eq018} reveals that the term $h_{m}$ determines the convergence rate of the algorithm, which is mainly influenced by the step-size. The mean-square convergence of the algorithm is ensured only if $|h_{m}|<1$. From this, the step-size should satisfy
\begin{equation}
	0<\mu<\frac{2L}{L+2}.
	\label{eq020}
\end{equation}

Therefore, the step-size must simultaneously satisfy conditions \eqref{eq012} and \eqref{eq020}, i.e., \mbox{$\mu \in \left( 0, \, \min \left\{ \frac{2}{\lambda_{\max}(\bm{\Lambda})}, \, \frac{2L}{L+2} \right\} \right)$} to guarantee the convergence of the algorithm, where $\text{min}\left\{\cdot\right\}$ represents to take the minimum value.

Moreover, when $h_{m}$ attains its minimum value, it holds
\begin{equation}
	\frac{\partial h_{m}}{\partial \mu}=2\mu\frac{L+2}{L^{2}}-2\frac{1}{L}=0.
	\label{eq021}
\end{equation}
At this point, the algorithm can reach the fastest convergence rate and the optimal step-size is $\mu_{\text{opt}}=\frac{L}{L+2}$ from \eqref{eq021}.
Taking \eqref{eq020} and $\mu_{\text{opt}}$ into account, we can draw the conclusion that in the range of $\mu \in (0, \mu_{\text{opt}}]$, the FxNLMS algorithm can converge and a larger step-size can lead to a faster convergence rate.
When the algorithm have gone into the steady-state, i.e., $J_{m+1}=J_{m}$, $m\rightarrow\infty$, we can get the steady-state MSD of the FxNLMS algorithm from \eqref{eq018}, i.e., 
\begin{equation}
	\displaystyle
		J(\infty)=\frac{\mu\sigma_{e}^{2}}{\sigma_{f}^{2}[2L-\mu(L+2)]}.
	\label{eq022}
\end{equation}
This equation signifies that the FxNLMS algorithm has a larger steady-state residual error for a larger step-size. The above properties will inspire us to propose the SSS-FxNLMS algorithm.

\vspace{0.4cm}
\subsection{SSS-FxNLMS algorithm}
We select a descending step-size sequence of length $K$, $\bm{\mu}=[\mu_{1},\mu_{2},\cdots,\mu_{K}]$, according to a rule that can be chosen based on the specific application scenario. By selecting an optimal step-size $\mu_{\text{opt},m}$ from $\bm{\mu}$ to update the FxNLMS weight vector at each iteration, the weight update formula of the SSS-FxNLMS algorithm can be formulated as
\begin{equation}
	\textbf{w}_{m+1}=\textbf{w}_{m}+\frac{\mu_{\text{opt},m}\textbf{x}_{f,m}e_{m}}{||\textbf{x}_{f,m}||_{2}^{2}}.
	\label{eq023}
\end{equation}

Below, we will illustrate how to get the optimal step-size for this iteration through predicting the MSD of the next iteration. For each step-size given in $\bm{\mu}$, the corresponding MSD trend will be calculated according to $\eqref{eq016}$ and $\eqref{eq017}$:
\begin{equation}
	J_{k,m+1}=\text{sum}[\overline{\textbf{P}}_{k,m+1}],
	\label{eq024}
\end{equation}
where $k=1,2,\cdots,K$, and
\begin{equation}
	\begin{aligned}
		\overline{\textbf{P}}_{k,m+1}=&(1-2\mu_{k}\frac{\overline{\textbf{R}}_{f,m}}{||\textbf{x}_{f,m}||_{2}^{2}})\odot\overline{\textbf{P}}_{k,m}+\mu^{2}_{k}\frac{\sigma_{e}^{2}\overline{\textbf{R}}_{f,m}}{||\textbf{x}_{f,m}||_{2}^{4}}\\
		&+2\mu^{2}_{k}\frac{\overline{\textbf{R}}_{f,m}\odot\overline{\textbf{P}}_{k,m}\odot\overline{\textbf{R}}_{f,m}}{||\textbf{x}_{f,m}||_{2}^{4}}\\
		&+\mu^{2}_{k}\frac{\overline{\textbf{R}}_{f,m}\text{sum}[\overline{\textbf{R}}_{f,m}\odot\overline{\textbf{P}}_{k,m}]}{||\textbf{x}_{f,m}||_{2}^{4}}.
		\label{eq025}
	\end{aligned}
\end{equation}
%We define $\textbf{J}_{m+1}=[J_{1,m+1},J_{2,m+1},\cdots,J_{K,m+1}]$ to collect MSD trends corresponding to different step-sizes.

Subsequently, we make a comparison of MSD trends associated with different step-sizes at iteration $m$ and then select the step-size corresponding to the minimum MSD trend as the optimal step-size $\mu_{\text{opt},m}$. For clarity, the process of selecting the optimal step size is formulated as
\begin{equation}
	\mu_{\text{opt},m}
	=
	\operatorname*{arg\,min}_{\mu_k}
	\left\{
	J_{1,m+1}, J_{2,m+1}, \cdots, J_{K,m+1}
	\right\}.
	\label{eq026}
\end{equation}
We set the initial value of the MSD trend to ${J}_{k,0}=\rho$. The initial value does not need to be highly accurate; this is due to the fact that the choice of the step-size depends on the trend of the MSD rather than its exact value. Extensive simulations reveal that the proposed SSS-based algorithm achieves good performance when $\rho\in[0.1,100]$, as visually confirmed in Fig. \ref{Fig2} in the simulation section.

\vspace{0.4cm}

\begin{table*}[t!]
	\centering
	\caption{Computational complexity for algorithms per iteration}
	\label{tab1}
	\setlength{\tabcolsep}{3pt} 
	\begin{tabular}{@{}ccccccc@{}}
		\hline
		Algorithms & Multiplications & Divisions & Additions & Other operators & {Run time (ms)} \\
		\hline
		{FxNLMS} & {$3L+M+1$} & {1} & {$3L+M-3$} & {-} &{0.003} \\
		{VSS1-FxNLMS \cite{kar2024improved}} & {$3L+M+4$} & {1} &{$3L+2M-2$} & {-} &{0.003} \\
		{VSS2-FxNLMS\;\, \cite{gomathi2016variable}} & {$3L+M+2$} & {1} &{$3L+M-3$} & {1} ${\text{exp}(\cdot)}$ {and} {1 $\text{arctan}(\cdot)$}& {0.003}\\
		{CC-FxNLMS \;\;\;\,\cite{song2019filtered}}& {$5L+2M+8$} & {2} &{$5L+2M$} & {1 $\text{exp}(\cdot)$} & {0.004}\\
		{SSS-FxNLMS} &{$6L+4KL+M+5$} & {yellow}{5K} & {$4L+4KL+M-4$} & {-} & {0.006}\\
		\hline
	\end{tabular}
	%	}
	\end{table*}
	\textbf{Remark 2:} Table 1 summarizes the computational complexity  of the FxNLMS, \mbox{VSS-FxNLMS}\footnote{It is obtained by using the VSS strategy from \cite{kar2024improved} and applying it to a single-channel feedforward ANC system.}\cite{kar2024improved},  \mbox{VSS-FxNLMS}\footnote{We utilize the VSS method in \cite{gomathi2016variable} for the FxNLMS algorithm to obtain it.}\cite{gomathi2016variable}, CC-FxCNLMS\footnote{By adopting the CC idea from \cite{song2019filtered}, we get the CC-FxNLMS algorithm.}\cite{song2019filtered}, and proposed SSS-FxNLMS algorithms in terms of the amount of arithmetic operations per iteration. Note that, to distinguish \mbox{VSS-based} FxNLMS algorithms in \cite{kar2024improved} and \cite{gomathi2016variable}, we use \mbox{VSS1-FxNLMS} and VSS2-FxNLMS to denote them respectively. As can be seen in Table 1, compared to the traditional FxNLMS algorithm, all the algorithms based on the time-varying step-size have increased computational complexity due to the integration of step-size calculation strategies. The complexity of the proposed SSS-FxNLMS algorithm is slightly higher than that of other algorithms, caused mainly by the MSD trend calculation related to preselected step-sizes. Importantly, the designed SSS scheme achieves improvements in both the convergence and steady-state behaviors for the FxNLMS algorithm. In addition, Table 1 provides the run time of these algorithms for each iteration on the laptop equipped with a 64-bit Windows operating system and an Intel® Core™ i7-9750H processor, where this processor has a main frequency of 2.60 GHz and adopts a 6-core 12-thread design.  It shows that the proposed SSS-FxNLMS algorithm has a run time within the same order of magnitude as the FxNLMS algorithm, still satisfying the current hardware operation requirements.
	
	\vspace{0.4cm}
	\textbf{Remark 3:} As stated in subsection \ref*{sec:Performance analysis of FxNLMS}, theoretically, the increase in the parameter $K$ indicates that the proposed \mbox{SSS-FxNLMS} algorithm can obtain a smooth convergence performance and a lower steady-state residual. However, as shown in Remark 2, the increase of $K$ also implies an increase in the computational complexity of the algorithm. Hence, inspired by this trade-off, we will introduce a sliding window dynamic step-size mechanism in the future work, which reserves two candidate step-sizes at each iteration to ensure performance while reducing the computational complexity of the algorithm. In particular, a maximum step-size is first selected from a pre-defined step-size sequence, and a smaller step-size is generated from it according to a certain ratio less than one, which serves as a candidate step-size for the next iteration. The MSD corresponding to the larger and smaller \mbox{step-sizes} is compared during each iteration. When the smaller step-size is consecutively selected for a preset number of iterations, an even smaller step-size is generated proportionally to replace the last larger \mbox{step-size}, thereby achieving dynamic adjustment of the step-size.
	\vspace{0.4cm}
	
	\subsection{R-SSS-FxNLMS algorithm}
	To enhance the robustness of the SSS-FxNLMS algorithm, we introduce a scaling factor $g[e_{m}]$ for it and modify the weight update formula $\left(\ref{eq005}\right)$ as
	\begin{equation}
\textbf{w}_{m+1}=\textbf{w}_{m}+\frac{\mu\textbf{x}_{f,m}g[e_{m}]}{||\textbf{x}_{f,m}||_{2}^{2}},
\label{eq027}
\end{equation}
where $g[e_{m}]\triangleq\frac{\varphi^{'}[e_{m}]}{e_{m}}$ and $\varphi^{'}[e_{m}]\triangleq\frac{\partial \varphi[e_{m}]}{\partial e_{m}}$, with $\varphi[e_{m}]$ being a robust function. Undoubtedly,  different robust functions in the existing literature can be used, with each generating a relevant algorithm.

Here, as an example of enhancing the robustness of the \mbox{SSS-FxNLMS} algorithm, we employ MCC for constructing the robust function $\varphi[e_{m}]$. The correntropy criterion is a statistical measure that represents the correlation between two random variables $C$ and $D$, which is defined as\cite{zhu2020robust}
\begin{equation}
V_{\sigma}(C,D)=\text{E}\left\{\kappa_{\sigma}(C,D)\right\},
\label{eq028}
\end{equation}
where $\kappa_{\sigma}(C,D)$ represents the kernel function. 

One of the commonly used types in correntropy is the Gaussian kernel, which is represented as
\begin{equation}
\varphi[e]=\kappa_{\sigma}=\text{exp}[-\frac{e^{2}}{2\sigma^{2}}],
\label{eq029}
\end{equation}
where $e=C-D$ and $\sigma>0$ represents the kernel width. By applying the MCC to the FxNLMS algorithm, we can obtain the expression for $g[e_{m}]$ in \eqref{eq027} as
\begin{equation}
g[e_{m}]=\text{exp}[-\frac{e^{2}_{m}}{2\sigma^{2}}].
\label{eq030}
\end{equation}

\vspace{0.4cm}

\textbf{Remark 4 :} We can draw the following conclusions from \eqref{eq027} and \eqref{eq030}. In the case of normal noise signals, the scaling factor $g[e_{m}]$ approaches 1, thus the performances of R-FxNLMS and FxNLMS are comparable, with \eqref{eq027} being equivalent to \eqref{eq005}. When impulsive noise emerges, $g[e_{m}]$ will become close to 0, which makes the algorithm update slowly, thereby weakening the influence of impulsive noise on the algorithm performance.

\vspace{0.4cm}

Subtracting both sides of $\left(\ref{eq027}\right)$ by $\textbf{w}_{o}$, yields

\begin{equation}
\widetilde{\textbf{w}}_{m+1}=\widetilde{\textbf{w}}_{m}-\frac{\mu\textbf{x}_{f,m}g[e_{m}]e_{m}}{||\textbf{x}_{f,m}||_{2}^{2}}.
\label{eq031}
\end{equation}

As both sides of $\left(\ref{eq031}\right)$ are post-multiplied by its transpose and the expectations of all the terms taken, we obtain
\begin{equation}
\begin{aligned}
	\textbf{P}_{m+1}=&\textbf{P}_{m}-\mu\text{E}\left\{\frac{\textbf{x}_{f,m}g[e_{m}]e_{m}}{||\textbf{x}_{f,m}||_{2}^{2}}\widetilde{\textbf{w}}^{\text{T}}_{m}\right\}\\
	&-\mu\text{E}\left\{\widetilde{\textbf{w}}_{m}\frac{\textbf{x}_{f,m}^{\text{T}}g[e_{m}]e_{m}}{||\textbf{x}_{f,m}||_{2}^{2}}\right\}\\
	&+\mu^{2}\text{E}\left\{g^2[e_{m}]e^{2}_{m}\frac{\textbf{x}_{f,m}\textbf{x}_{f,m}^{\text{T}}}{||\textbf{x}_{f,m}||_{2}^{4}}\right\}.
	\label{eq032}
\end{aligned}
\end{equation}

Then, by performing simplification operations similar to those in Appendix A and vectorization operations similar to those in \eqref{eq016}, $\left(\ref{eq032}\right)$ can be expressed as
\begin{equation}
\begin{aligned}
	\overline{\textbf{P}}_{m+1}=&(1-2\mu g[e_{m}]\frac{\overline{\textbf{R}}_{f,m}}{||\textbf{x}_{f,m}||_{2}^{2}})\odot\overline{\textbf{P}}_{m}+\mu^{2}g^{2}[e_{m}]\frac{\sigma_{e}^{2}\overline{\textbf{R}}_{f,m}}{||\textbf{x}_{f,m}||_{2}^{4}}\\
	&+2\mu^{2}g^{2}[e_{m}]\frac{\overline{\textbf{R}}_{f,m}\odot\overline{\textbf{P}}_{m}\odot\overline{\textbf{R}}_{f,m}}{||\textbf{x}_{f,m}||_{2}^{4}}\\
	&+\mu^{2}g^{2}[e_{m}]\frac{\overline{\textbf{R}}_{f,m}\text{sum}[\overline{\textbf{R}}_{f,m}\odot\overline{\textbf{P}}_{m}]}{||\textbf{x}_{f,m}||_{2}^{4}}.
	\label{eq033}
\end{aligned}
\end{equation}

Together with \eqref{eq017}, \eqref{eq033} can model the MSD trend of the \mbox{SSS-FxNLMS} algorithm. With the derived MSD recursion, we can directly take advantage of the SSS scheme in the previous subsection to develop the R-SSS-FxNLMS algorithm.

For clarify, we summarize the implementation process of these SSS-FxNLMS and \mbox{R-SSS-FxNLMS} algorithms in \mbox{Table 2}.

\begin{table}[htbp]
	\caption{Summary of the SSS-FxNLMS and R-SSS-FxNLMS algorithms}
	\label{table}
	\setlength{\tabcolsep}{3pt}
%	\resizebox{\linewidth}{!}{
	\small{
		\begin{tabular}{l l}
			\hline
			Parameters:  $\rho$, $K$, $\bm{\mu}$, $\sigma$\\
			\hline
			Initialization: $\textbf{J}_{1}=\rho{\textbf{I}_{1\times{K}}}$, $\textbf{w}(0)=[\textbf{0}_{1\times{L}}]$ \\
			\hline
			for	$m$=$1$,2,3,$\cdots$\\
			\;\;\;\;\;\;(a) To calculate the output signal $y_{m}$ of the adaptive filter\\
			\;\;\;\;\;\;\;\;\;\;\;\;$y_{m}=\textbf{x}^{\text{T}}_{m}\textbf{w}_{m}$\\
			\;\;\;\;\;\;(b) To calculate the secondary cancelling signal $\hat{d}_{m}$\\
			\;\;\;\;\;\;\;\;\;\;\;\;$\hat{d}_{m}=\textbf{y}^{\text{T}}_{m}\textbf{s}_{m}$.\\
			\;\;\;\;\;\;(c) To calculate the error signal $e_{m}$\\
			\;\;\;\;\;\;\;\;\;\;\;\;$e_{m}=d_{m}-\hat{d}_{m}$\\
			\;\;\;\;\;\;(d) To calculate $\sigma^{2}_{e,m}$ and $\overline{\textbf{R}}_{f,m}$\\
			\;\;\;\;\;\;\;\;\;\;\;\;$\sigma^{2}_{e,m}=\lambda\sigma^{2}_{e,m}+(1-\lambda){e}^{2}_{m}$\\
			\;\;\;\;\;\;\;\;\;\;\;\;$\overline{\textbf{R}}_{f,m}=\textbf{x}_{f,m}\odot\textbf{x}_{f,m}$\\
			\;\;\;\;\;\;(e) To calculate the MSD corresponding to different step-sizes \\
			\;\;\;\;\;\;\;\;\;\;\;\;$g[e_{m}]=\left\{ \begin{aligned}
				&1\quad\quad\quad\quad, \mbox{\bfseries SSS-FxNLMS algorithm}\\
				&\text{exp}[-\frac{e^{2}_{m}}{2\sigma^{2}}], \mbox{\bfseries R-SSS-FxNLMS algorithm}\\
			\end{aligned} \right.$\\
			\;\;\;\;\;\;\;\;\;\;\;\;for k=1:K\\
			\;\;\;\;\;\;\;\;\;\;\;\;\;\;\;\;	
			$\overline{\textbf{P}}_{k,m+1}=(1-2\mu g[e_{m}]\frac{\overline{\textbf{R}}_{f,m}}{||\textbf{x}_{f,m}||_{2}^{2}})\odot\overline{\textbf{P}}_{k,m}$\\
			\;\;\;\;\;\;\;\;\;\;\;\;\;\;\;\;\;\;\;\;\;\;\;\;\;\;\;\;\;\;\;\;\;\;$+\mu^{2}g^{2}[e_{m}]\frac{\sigma_{e}^2\overline{\textbf{R}}_{f,m}}{||\textbf{x}_{f,m}||_{2}^{4}}$\\
			\;\;\;\;\;\;\;\;\;\;\;\;\;\;\;\;\;\;\;\;\;\;\;\;\;\;\;\;\;\;\;\;\;\;$+2\mu^{2}g^{2}[e_{m}]\frac{\overline{\textbf{R}}_{f,m}\odot\overline{\textbf{P}}_{k,m}\odot\overline{\textbf{R}}_{f,m}}{||\textbf{x}_{f,m}||_{2}^{4}}$\\
			\;\;\;\;\;\;\;\;\;\;\;\;\;\;\;\;\;\;\;\;\;\;\;\;\;\;\;\;\;\;\;\;\;\;$+\mu^{2}g^{2}[e_{m}]\frac{\overline{\textbf{R}}_{f,m}\text{sum}[\overline{\textbf{R}}_{f,m}\odot\overline{\textbf{P}}_{k,m}]}{||\textbf{x}_{f,m}||_{2}^{4}}$\\
			\;\;\;\;\;\;\;\;\;\;\;\;\;\;\;\;$J_{k,m}=\text{sum}[\overline{\textbf{P}}_{k,m}]$\\
			\;\;\;\;\;\;\;\;\;\;\;\;end\\	
			\;\;\;\;\;\;\;\;\;\;\;\;$\textbf{J}_{m+1}=[J_{1,m+1}, J_{2,m+1},\cdots,J_{K,m+1}]$\\	
			\;\;\;\;\;\;(f) To choose the optimal step-size\\
			\;\;\;\;\;\;\;\;\;\;\;\;$\mu_{\text{opt},m}=\mathop{\arg\min}\limits_{\mu_{k}} \left\{\textbf{J}_{m+1}\right\}$\\
			\;\;\;\;\;\;(g) To update the weight vector\\
			\;\;\;\;\;\;\;\;\;\;\;\;$\textbf{w}_{m+1}=\textbf{w}_{m}+\frac{\mu_{\text{opt},m}\textbf{x}_{f,m}g[e_{m}]e_{m}}{||\textbf{x}_{f,m}||_{2}^{2}}$ \\
			end\\
			\hline
	\end{tabular}}
	\label{tab2}
\end{table}

\section{Simulation Results}

In this section, we conducte experiments using three common types of noise as reference inputs to validate the performance of the proposed algorithms. The transfer function of the primary path is\cite{zhang2019normalized}
\begin{equation}
	\begin{aligned}
		P(z)=&0.01-0.05z^{-1}+0.02z^{-2}+0.75z^{-3}-0.4z^{-4}\\
		&-0.5z^{-5}-0.2z^{-6}-0.05z^{-7}+0.3z^{-8}+0.005z^{-9}.
		\label{eq034}
	\end{aligned}
\end{equation}
The transfer function of the secondary path is
\begin{equation}
	\begin{aligned}
		S(z)=0.01-0.01z^{-1}+0.9z^{-2}+0.02z^{-3}-0.5z^{-4}.
		\label{eq035}
	\end{aligned}
\end{equation}
In the following experiments, it is assumed that $\hat{S}(z)=S(z)$ to obtain the filtered reference input $x_{f,m}$, unless otherwise specified.
The adaptive filter is set to have a length of $L=16$. The averaged noise reduction (ANR) is used to evaluate the performance of the algorithm, formulated as
\begin{equation}
	\text{ANR}_{m}=20\text{log}_{10}(\frac{A_{e,m}}{A_{d,m}}),
	\label{eq036}
\end{equation}
where $A_{e,m}=\beta A_{e,m-1}+(1-\beta)|e_{m}|$ and $A_{d,m}=\beta A_{d,m-1}+(1-\beta)|d_{m}|$, with $\beta=0.999$ being the forgetting factor. All simulation results are obtained by averaging 100 trials unless otherwise specified. The parameters selected for each algorithm are listed in detail in Tables 3.

\begin{table*}[t!]
	\centering
	\caption{Algorithm parameters under different noise conditions}
	\label{tab3}
	\setlength{\tabcolsep}{4pt}
	\begin{tabular}{@{}ccc@{}}
		\hline
		\textbf{Algorithms}
		& \textbf{Parameters}
		&  \\ 
		\hline
		
		% ================= Gaussian white noise =================
		\multicolumn{3}{c}{\textbf{Gaussian white noise (Figs. 2--8)}} \\
		\hline
		SSS-FxNLMS
		& $K=4$, $\bm{\mu}=[0.6, 0.3, 0.15, 0.075]$, $\lambda=0.8$
		&  \\
		VSS1-FxNLMS  \cite{kar2024improved} 
		& $\mu_{upperbound}=0.6$, $\mu_{lowerbound}=0.075$, $\alpha=0.97$, $\gamma=5.5\times10^{-2}$
		&  \\
		VSS2-FxNLMS   \cite{gomathi2016variable} 
		& $\beta=0.6$, $\gamma=1$
		&  \\
		CC-FxNLMS  \cite{song2019filtered} 
		& $\mu_{1}=0.6$, $\mu_{2}=0.075$, $\mu_{a}=25$
		&  \\
		\hline
		
		% ================= Gaussian colored noise =================
		\multicolumn{3}{c}{\textbf{Gaussian colored noise (Figs. 10--12)}} \\
		\hline
		{SSS-FxNLMS} 
		& {$K=4$, $\bm{\mu}=[0.36, 0.18, 0.09, 0.045]$, $\lambda=0.8$} 
		&  \\
		{AD-SSS-FxNLMS} 
		& {$K=4$, $\bm{\mu}=[0.36, 0.18, 0.09, 0.045]$, $\lambda=0.8$} 
		&  \\
		{AD-VSS1-FxNLMS}  \cite{kar2024improved} 
		& {$\mu_{upperbound}=0.36$, $\mu_{lowerbound}=0.045$, $\alpha=0.97$, $\gamma=5.5\times10^{-2}$} 
		&  \\
		{AD-VSS2-FxNLMS }  \cite{gomathi2016variable} 
		& {$\beta=0.36$, $\gamma=1$} 
		&  \\
		{AD-CC-FxNLMS }  \cite{song2019filtered} 
		& {$\mu_{1}=0.36$, $\mu_{2}=0.045$, $\mu_{a}=25$} 
		&  \\
		\hline
		
		% ================= alpha-stable noise =================
		\multicolumn{3}{c}{$\alpha$\textbf{-stable noise (Figs. 13--15, 17)}} \\
		\hline
		{R-SSS-FxNLMS (MCC-based)} 
		& {$K=4$, $\bm{\mu}=[0.6, 0.3, 0.15, 0.075]$, $\lambda=0.8$} 
		& \\
	{R-VSS1-FxNLMS (MCC-based)}   \cite{kar2024improved} 
		& {$\mu_{upperbound}=0.6$, $\mu_{lowerbound}=0.075$, $\alpha=0.97$, $\gamma=5.5\times10^{-2}$} 
		&  \\
		{R-VSS2-FxNLMS (MCC-based) } \cite{gomathi2016variable} 
		& {$\beta=0.6$, $\gamma=1$} 
		&  \\
		{R-CC-FxNLMS (MCC-based) } \cite{song2019filtered} 
		& {$\mu_{1}=0.6$, $\mu_{2}=0.075$, $\mu_{a}=25$} 
		&  \\
		\hline
		
		% ================= alpha-stable noise =================
		\multicolumn{3}{c}{$\alpha$\textbf{-stable noise (Figs. 16)}} \\
		\hline
		{R-SSS-FxNLMS (EHCF-based)} 
		& {$K=4$, $\bm{\mu}=[0.15,0.075,0.0375,0.01875]$, $\lambda=0.8$} 
		&  \\
		{R-VSS1-FxNLMS (EHCF-based) }  \cite{kar2024improved} 
		& {$\mu_{upperbound}=0.6$, $\mu_{lowerbound}=0.075$, $\alpha=0.97$, $\gamma=5.5\times10^{-2}$} 
		&  \\
		{R-VSS2-FxNLMS (EHCF-based) } \cite{gomathi2016variable} 
		& {$\beta=0.6$, $\gamma=1$} 
		&  \\
	{R-CC-FxNLMS (EHCF-based) } \cite{song2019filtered} 
		& {$\mu_{1}=0.6$, $\mu_{2}=0.075$, $\mu_{a}=25$} 
		&  \\
		\hline
		
		% ================= Factory noise =================
		\multicolumn{3}{c}{\textbf{Factory noise (Figs. 19-20)}} \\
		\hline
		{R-SSS-FxNLMS (MCC-based)} 
		& {$K=3$, $\bm{\mu}=[0.4, 0.24, 0.144]$, $\lambda=0.8$} 
		&  \\
		{R-VSS1-FxNLMS (MCC-based) }  \cite{kar2024improved} 
		& {$\mu_{upperbound}=0.4$, $\mu_{lowerbound}=0.144$, $\alpha=0.97$, $\gamma=5.5\times10^{-2}$} 
		&  \\
		{R-VSS2-FxNLMS (MCC-based) }  \cite{gomathi2016variable} 
		& {$\beta=0.36$, $\gamma=0.3$} 
		&  \\
		{R-CC-FxNLMS (MCC-based) }  \cite{song2019filtered} 
		& {$\mu_{1}=0.36$, $\mu_{2}=0.045$, $\mu_{a}=25$} 
		&  \\
		\hline
		
		% ================= Hammering noise =================
		\multicolumn{3}{c}{\textbf{Hammering noise (Fig. 22)}} \\
		\hline
		{SSS-FxNLMS} 
		& {$K=3$, $\bm{\mu}=[0.4, 0.24, 0.144]$, $\lambda=0.8$} 
		&  \\
		{VSS1-FxNLMS }  \cite{kar2024improved} 
		& {$\mu_{upperbound}=0.4$, $\mu_{lowerbound}=0.144$, $\alpha=0.97$, $\gamma=5.5\times10^{-2}$} 
		&  \\
		{VSS2-FxNLMS }  \cite{gomathi2016variable} 
		& {$\beta=0.2$, $\gamma=1$} 
		&  \\
	{CC-FxNLMS }  \cite{song2019filtered} 
		& {$\beta=0.2$, $\gamma=1$ (VSS2-FxNLMS); $\mu_{1}=0.4$, $\mu_{2}=0.144$, $\mu_{a}=0.01$} 
		&  \\
		\hline
	\end{tabular}
\end{table*}

\subsection{Experiment 1: Gaussian noise}
In this subsection, we use the Gaussian white noise as the reference input signal except Figs. \ref{Fig10}-\ref{Fig12}. Based on extensive simulation tests, an equal ratio sequence is selected  as the step-sizes, i.e., $\mu_{k}=0.5\mu_{k-1}$.

(1) Selection of $\rho$

\begin{figure}[htbp]
	\centering
	{
		\begin{minipage}[a]{0.4\textwidth}
			\includegraphics[width=1\textwidth]{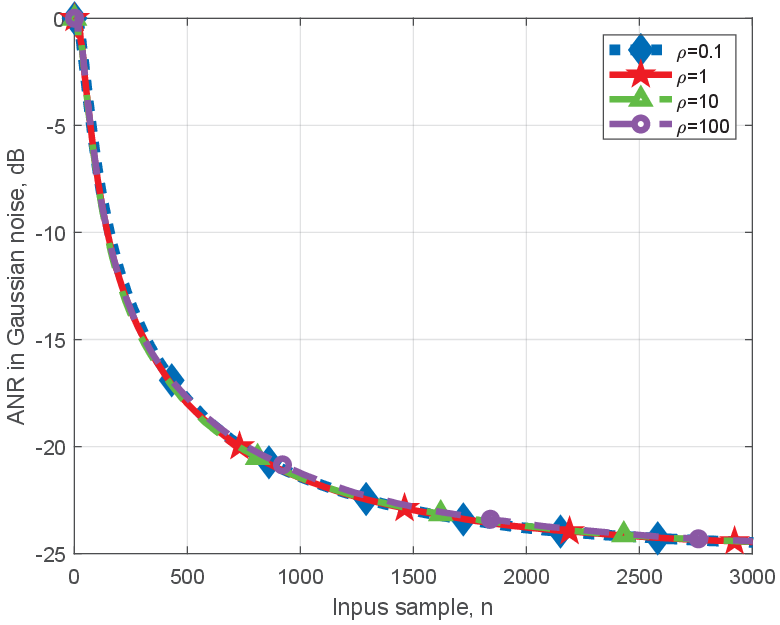} 
		\end{minipage}
	}
	\caption{ANR curves of the SSS-FxNLMS algorithm versus different $\rho$.}
	\label{Fig2}
\end{figure}

The effect of $\rho$ on the performance of the SSS-FxNLMS algorithm is shown in Fig. \ref{Fig2}. It can be observed that this algorithm exhibits good performance when $\rho\in[0.1,100]$, indicating that the initial value has little effect on the algorithm’s performance. In the following discussion, thus we set $\rho=1$.

(2) Comparison of algorithms in Gaussian white noise

\begin{figure}[htbp]
	\centering
	{
		\begin{minipage}[a]{0.4\textwidth}
			\includegraphics[width=1\textwidth]{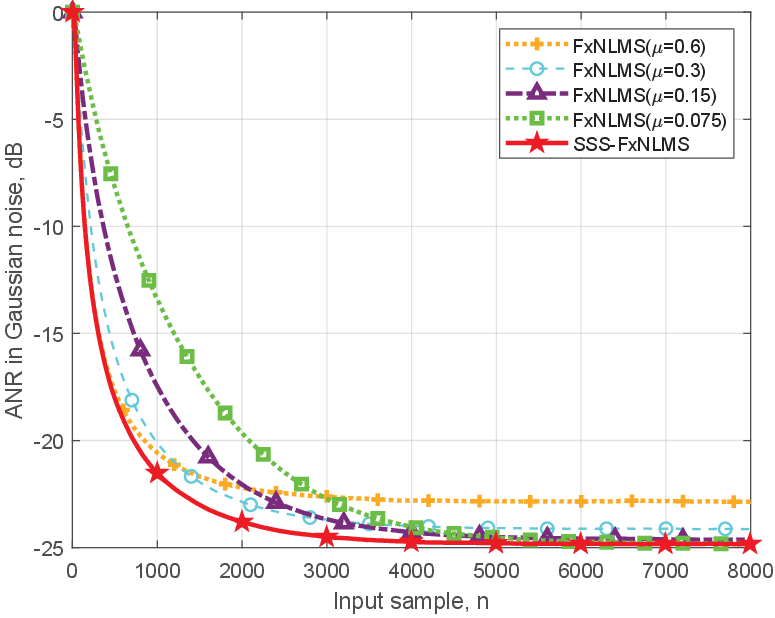} 
		\end{minipage}
	}
	\caption{ANR curves of the SSS-FxNLMS algorithm and standard FxNLMS algorithm with four fixed step-sizes in Gaussian white noise.}
	\label{Fig3}
\end{figure}

\begin{figure}[htbp]
	\centering
	{
		\begin{minipage}[a]{0.4\textwidth}
			\includegraphics[width=1\textwidth]{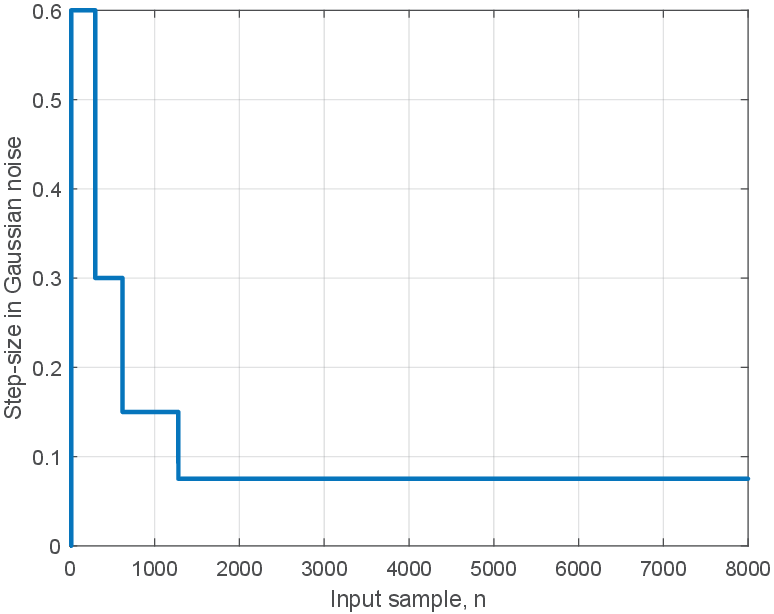} 
		\end{minipage}
	}
	\caption{The switching process of step-sizes in the SSS-FxNLMS algorithm.}
	\label{Fig4}
\end{figure}

Fig. \ref{Fig3} compares the ANR curves of the proposed  \mbox{SSS-FxNLMS} algorithm and FxNLMS algorithms with that of four different fixed step-size. As mentioned in the \mbox{Remark 1}, the FxNLMS algorithm with a large step-size demonstrates a fast convergence rate but an inferior steady-state residual error. A small step-size indicates a slow convergence rate, while it results a lower steady-state residual error. The proposed \mbox{SSS-FxNLMS} algorithm get fast convergence rate and low steady-state residual error simultaneously, owing to its ability to select the optimal step-size at various convergence stages. In order to further explain the features of the switched step-size, the step-size switching process is given in \mbox{Fig. \ref{Fig4}}.

\begin{figure}[htbp]
	\centering
	{
		\begin{minipage}[a]{0.4\textwidth}
			\includegraphics[width=1\textwidth]{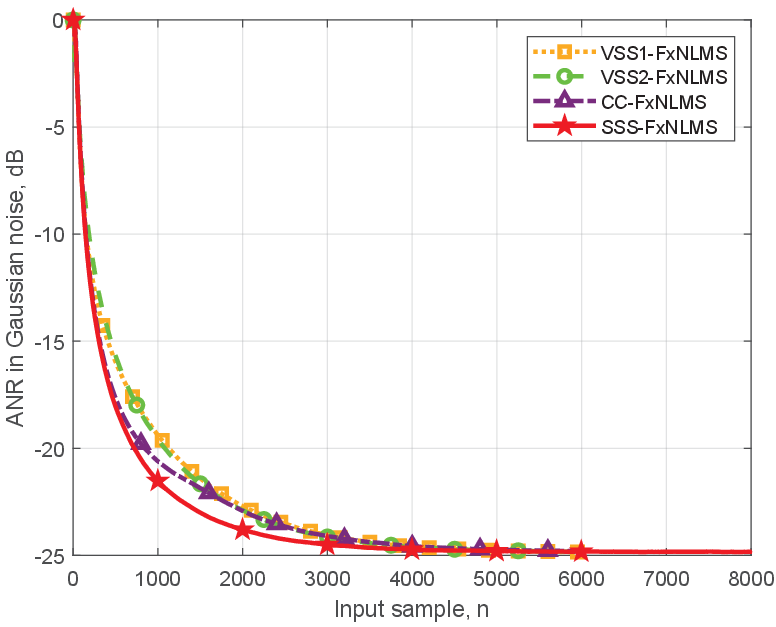} 
		\end{minipage}
	}
	\caption{ANR curves of different algorithms in Gaussian white noise.}
	\label{Fig5}
\end{figure}

In Fig. \ref{Fig5}, the ANR performance of proposed SSS-FxNLMS algorithm is compared with that of the \mbox{VSS1-FxNLMS\cite{kar2024improved}}, VSS2-FxNLMS\cite{gomathi2016variable} and CC-FxCNLMS\cite{song2019filtered} algorithms. According to the principle that all the algorithms share similar \mbox{steady-state} residual error, their respective parameters are chosen. In \mbox{Fig. \ref{Fig5}}, it can be seen that the \mbox{steady-state} residual error of the four algorithms are similar. The VSS1-FxNLMS and \mbox{VSS2-FxNLMS} algorithms exhibit approximate convergence rate. But the \mbox{CC-FxNLMS} algorithm outperforms both, and the proposed \mbox{SSS-FxNLMS} algorithm is superior to \mbox{CC-FxNLMS}. In addition, compared with the CC-FxNLMS algorithm, the proposed SSS-FxNLMS algorithm has the advantages of easy parameter selection and low computational complexity.
\begin{figure}[htbp]
	\centering
	{
		\begin{minipage}[a]{0.4\textwidth}
			\includegraphics[width=1\textwidth]{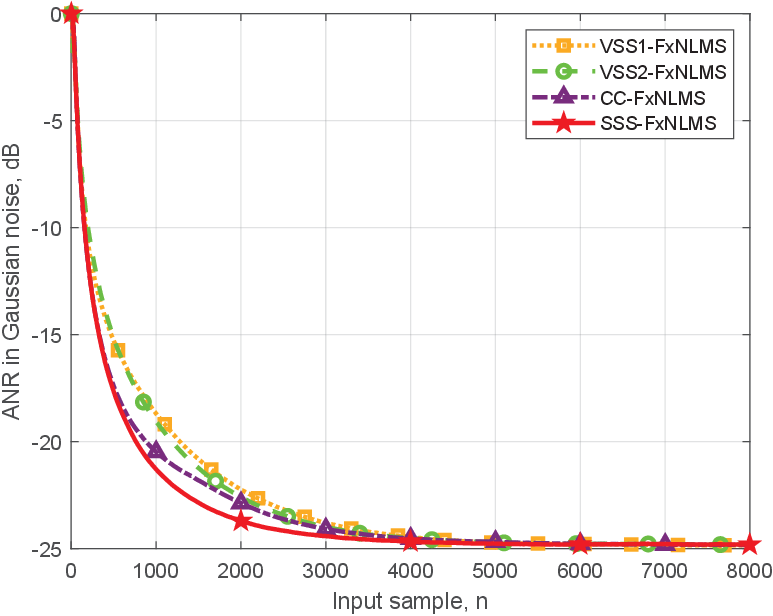} 
		\end{minipage}
	}
	\caption{ANR curves of different algorithms with $\hat{S}(z)$ in Gaussian white noise.}
	\label{Fig6}
\end{figure}

In practical ANC system, it would be difficult to model the secondary path $S(z)$ perfectly. Usually, the secondary path $S(z)$ needs to be estimated in advance. Hence, in this example, we use the estimated secondary path $\hat{S}(z) =-0.0455-$ $0.0453z^{-1}+0.8683z^{-2}+0.0399z^{-3}-0.518z^{-4}$ to obtain the filtered reference input vector $\bm x_{f,m}$ for all the compared ANC algorithms, where $\hat{S}(z)$ is obtained by using the Eriksson's classic identification method presented in \cite{eriksson1989use} and it has a error of $5.7\%$ compared with its accurate $S(z)$. The ANR results are depicted in Fig. \ref{Fig6}. It shows that the proposed SSS-FxNLMS algorithm is still superoir to its counterparts even in the case that the estimated secondary path is not perfect.

\begin{figure}[htbp]
	\centering
	{
		\begin{minipage}[a]{0.4\textwidth}
			\includegraphics[width=1\textwidth]{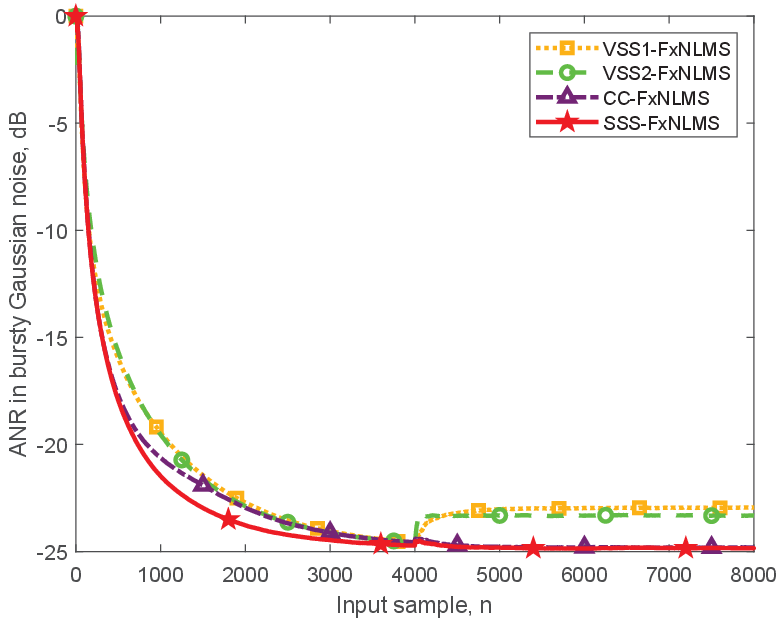} 
		\end{minipage}
	}
	\caption{ANR curves of different algorithms in bursty Gaussian white noise.}
		\label{Fig7}
\end{figure}

\begin{figure}[htbp]
	\centering
	{
		\begin{minipage}[a]{0.4\textwidth}
			\includegraphics[width=1\textwidth]{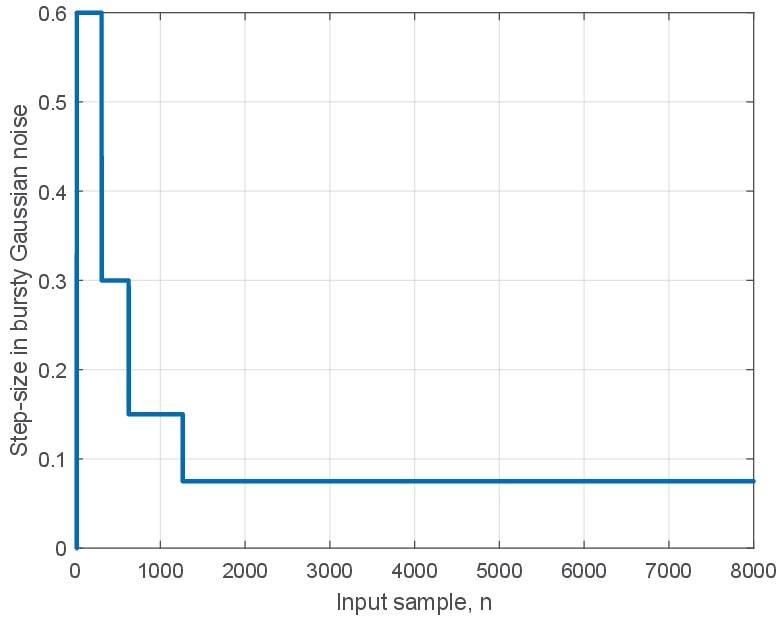} 
		\end{minipage}
	}
	\caption{The switching process of step-sizes in the SSS-FxNLMS algorithm.}
		\label{Fig8}
\end{figure}

To assess the adaptability of the proposed SSS-based algorithm under bursty noise conditions, we consider a piecewise Gaussian white noise sequence: the first half of the samples have zero mean and unit variance, followed by an abrupt increase in variance to 100 for the second half. The resulting ANR curves of the different algorithms in this scenario are shown in Fig. \ref{Fig7}, from which it can be observed that the SSS-FxNLMS algorithm exhibits superior performance. \mbox{Fig. \ref{Fig8}} depicts the step-size switching behavior of the proposed \mbox{SSS-FxNLMS} algorithm. It is worth noting that in feedforward ANC systems with fixed primary and secondary paths, the optimal filter weights remain constant. The role of the adaptive filter is therefore to gradually converge to this optimal solution, meaning that the switched step-size generally decreases over time as convergence progresses, rather than responding directly to instantaneous noise fluctuations.


(3) Comparison of algorithms in Gaussian colored noise

\begin{figure}[htbp]
	\centering
	{
		\begin{minipage}[a]{0.45\textwidth}
			\includegraphics[width=1\textwidth]{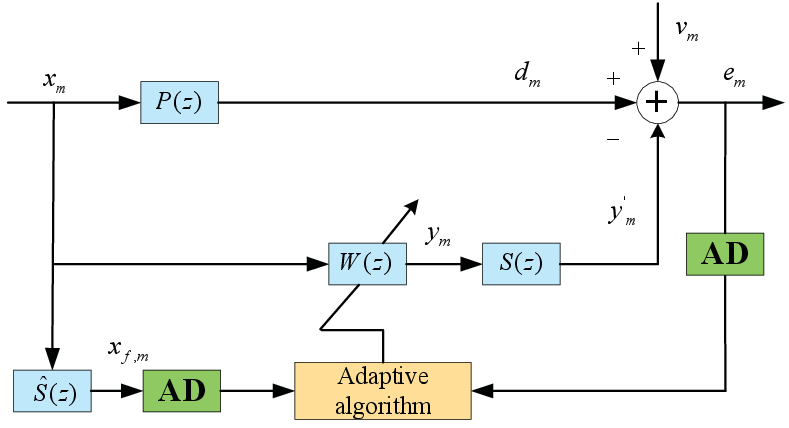} 
		\end{minipage}
	}
	\caption{The AD structure for the ANC system.}
	\label{Fig9}
\end{figure}

\begin{figure}[htbp]
	\centering
	{
		\begin{minipage}[a]{0.4\textwidth}
			\includegraphics[width=1\textwidth]{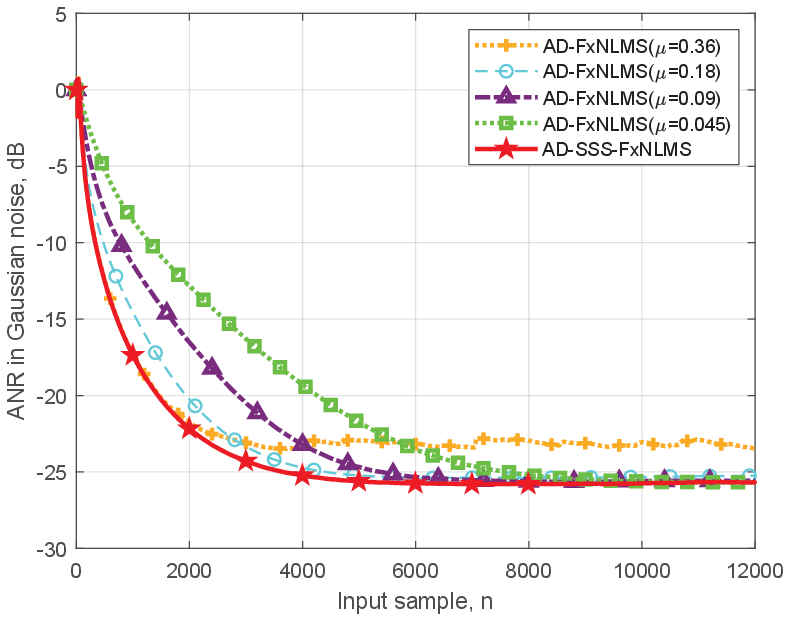} 
		\end{minipage}
	}
	\caption{ANR curves of the AD-SSS-FxNLMS and standard AD-FxNLMS algorithms in Gaussian colored noise environment.}
	\label{Fig10}
\end{figure}

\begin{figure}[htbp]
	\centering
	{
		\begin{minipage}[a]{0.4\textwidth}
			\includegraphics[width=1\textwidth]{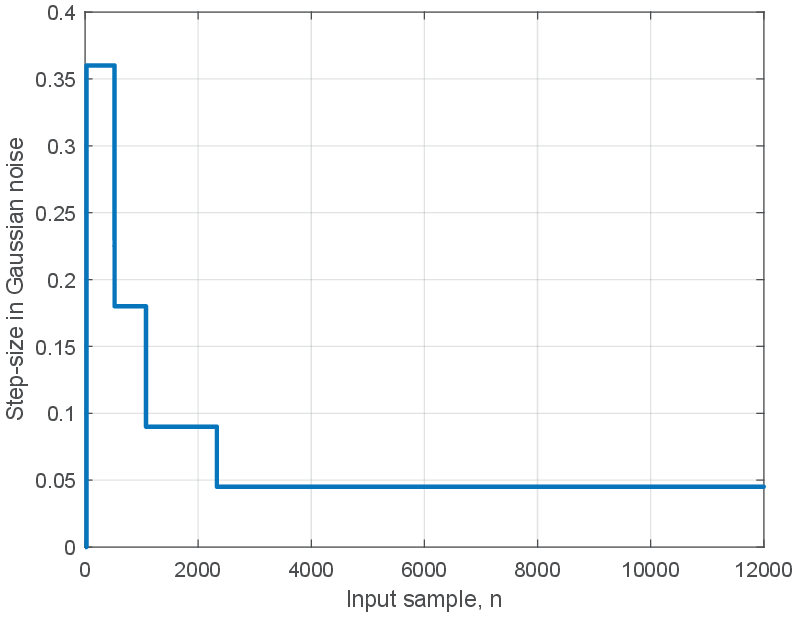} 
		\end{minipage}
	}
	\caption{The switching process of step-sizes in the AD-SSS-FxNLMS algorithm.}
	\label{Fig11}
\end{figure}

\begin{figure}[htbp]
	\centering
	{
		\begin{minipage}[a]{0.4\textwidth}
			\includegraphics[width=1\textwidth]{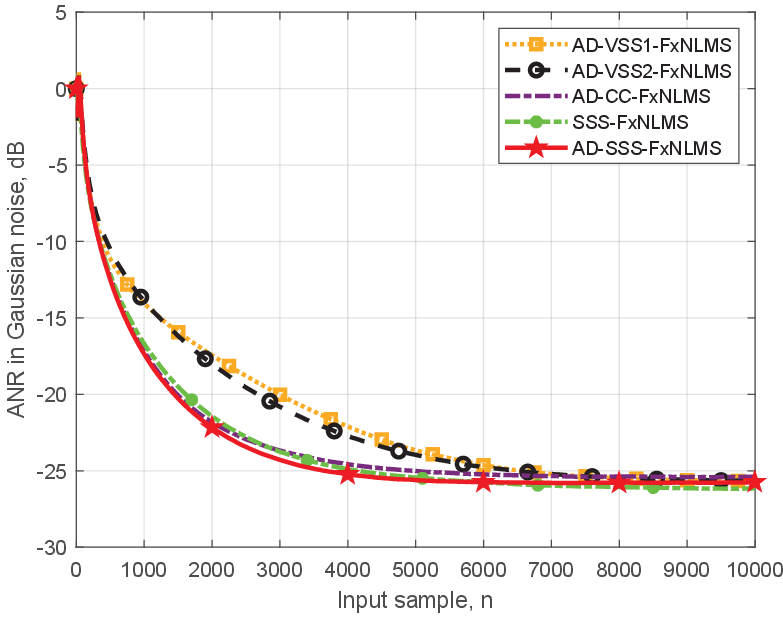} 
		\end{minipage}
	}
	\caption{ANR curves of different algorithms in Gaussian colored noise. The selection of parameters for algorithms is set to: $\mu_{upperbound}=0.36$, $\mu_{lowerbound}=0.045$, $\alpha=0.97$, $\gamma=5.5\times10^{-2}$ (AD-VSS1-FxNLMS); $\beta=0.36$, $\gamma=1$ (AD-VSS2-FxNLMS); $\mu_{1}=0.36$, $\mu_{2}=0.045$, $\mu_{a}=25$ (\mbox{AD-CC-FxNLMS}); $K=4$, $\bm{\mu}=[0.36, 0.18, 0.09, 0.045]$, $\lambda=0.8$ (SSS-FxNLMS). Parameters of AD-SSS-FxNLMS refer to Fig. \ref{Fig10}. }
	\label{Fig12}
\end{figure}

In this example, we explore the performance of the \mbox{SSS-FxNLMS} algorithm in colored noise environments. The colored noise signal $x(n)$ is generated by filtering a zero-mean and unit-variance white Gaussian sequence through the \mbox{first-order} AR system $F(z)={1}/{(1-0.9z^{-1})}$. Considering the weak decorrelation of NLMS-based algorithms for colored noises, we take advantage of the adaptive decorrelation (AD) technique in \cite{zhang2017family} to preprocess the colored noise signal. The AD structure for the ANC system is shown in Fig. \ref{Fig9}. As such, we obtain the AD-based versions for the algorithms in Figs. \ref{Fig10}-\ref{Fig12}, i.e., named as AD-FxNLMS, \mbox{AD-VSS1-FxNLMS}, AD-VSS2-FxNLMS, \mbox{AD-CC-FxNLMS}, and AD-SSS-FxNLMS, respectively. Fig. \ref{Fig10} compares the ANR curves of the AD-FxNLMS algrithm with fixed \mbox{step-sizes} and the AD-SSS-FxNLMS algrithm.  The adaptive decorrelation parameters in \cite{zhang2017family} are $K=10$ and $\mu_{a}=0.001$. As shown in Fig. \ref{Fig10}, the AD-SSS-FXNMS algorithm still achieves both fast convergence of large step-size and low steady-state residual error of small step-size, due to the fact that it utilizes the advantages of step-size switching, where one can see Fig. \ref{Fig11} for its step-size switching process. Fig. \ref{Fig12} shows the ANR curves of the AD-VSS1-FxNLMS, AD-VSS2-FxNLMS, AD-CC-FxNLMS, SSS-FxNLMS and \mbox{AD-SSS-FxNLMS} algorithms. Clearly, thanks to the use of the AD technique, the AD-SSS-FxNLMS algorithm converges faster than the SSS-FxNLMS algorithm. Among these AD-based FxNLMS algorithms for eliminating colored noise inputs, the proposed AD-SSS-FxNLMS algorithm also exhibits the best performance in the convergence rate.

\subsection{Experiment 2: $\alpha$-stable noise}

The impulsive noise is generated from the $\alpha$-stable model, whose characteristic function is given by\cite{yu2019m}
\begin{equation}
	\Phi(t)=\text{exp}(-\gamma|t|^{\alpha}),
	\label{eq037}
\end{equation}
where $\gamma$ is the coefficient that indicates the extent of noise dispersion, while $\alpha$ is the characteristic indicator for controlling the impulsiveness of the function. We choose $\alpha=1.4$ and $\gamma=0.1$. The step-size selection rule is the same as \mbox{Experiment 1}.

\begin{figure}[htbp]
	\centering
	{
		\begin{minipage}[a]{0.4\textwidth}
			\includegraphics[width=1\textwidth]{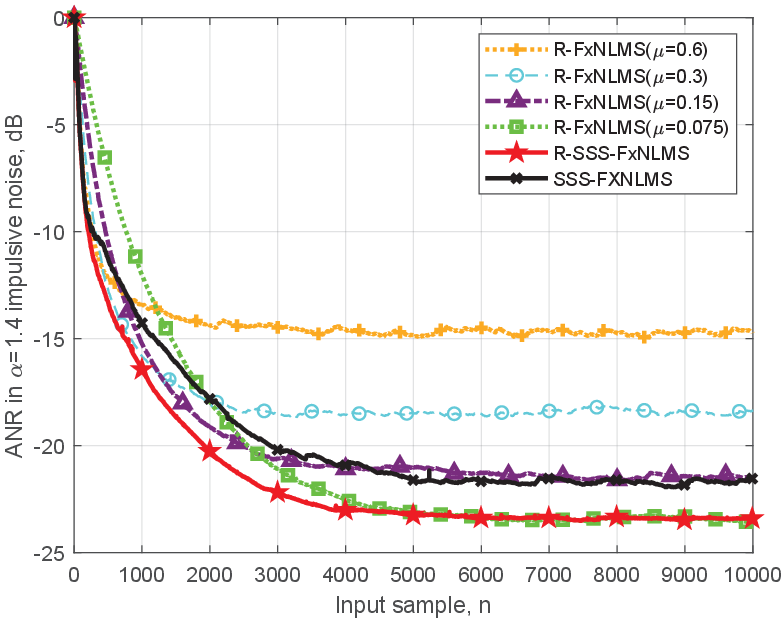} 
		\end{minipage}
	}
	\caption{ANR curves of the R-SSS-FxNLMS algorithm, \mbox{SSS-FxNLMS} algorithm and standard R-FxNLMS algorithms with four fixed \mbox{step-sizes} in $\alpha=1.4$ impulsive noise. The other parameters of the \mbox{R-SSS-FxNLMS}.}
	\label{Fig13}
\end{figure}

\begin{figure}[htbp]
	\centering
	{
		\begin{minipage}[a]{0.4\textwidth}
			\includegraphics[width=1\textwidth]{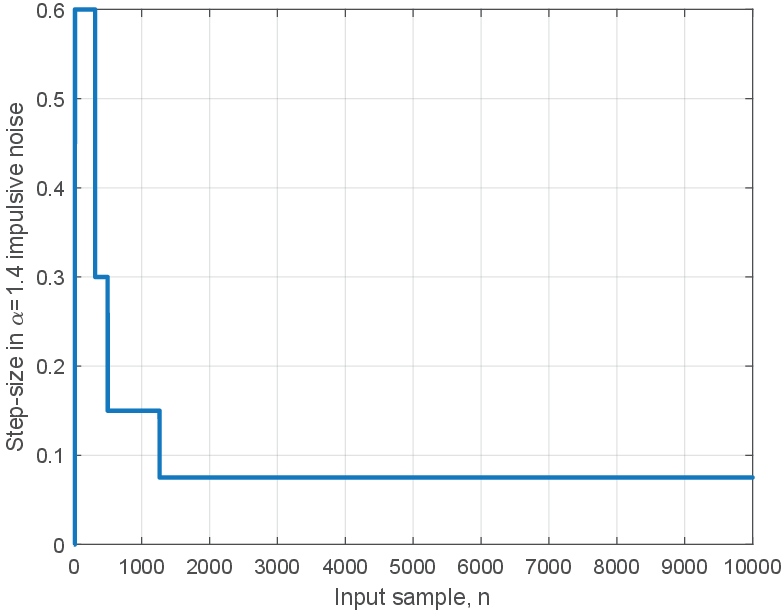} 
		\end{minipage}
	}
	\caption{The switching process of step-sizes in the R-SSS-FxNLMS algorithm.}
	\label{Fig14}
\end{figure}

\begin{figure}[htbp]
	\centering
	{
		\begin{minipage}[a]{0.4\textwidth}
			\includegraphics[width=1\textwidth]{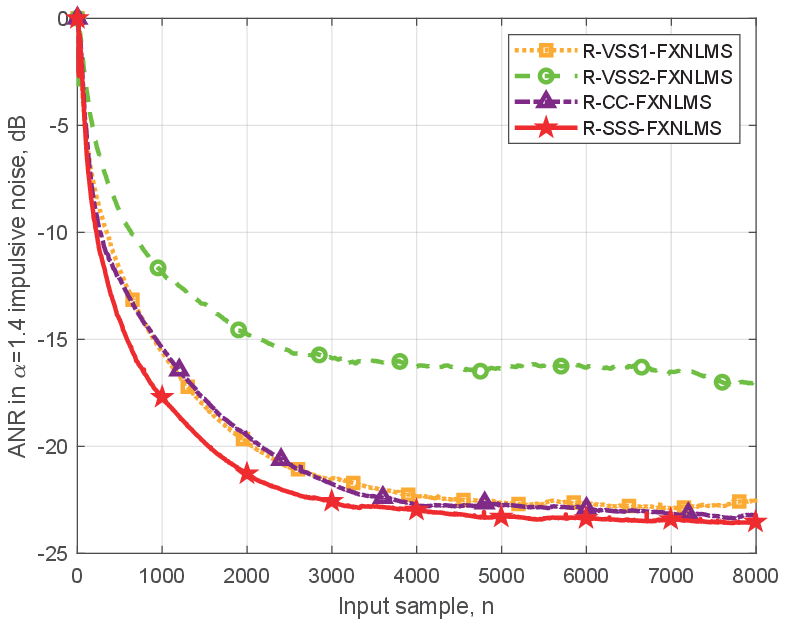} 
		\end{minipage}
	}
	\caption{ANR curves of different algorithms in $\alpha=1.4$ impulsive noise.}
	\label{Fig15}
\end{figure}

To offer a fair comparison, the MCC robust strategy is incorporated into the FxNLMS, \mbox{VSS1-FxNLMS}, \mbox{VSS2-FxNLMS} and \mbox{CC-FxCNLMS} algorithms, and then resulting in the \mbox{R-FxNLMS}, R-VSS1-FxNLMS, R-VSS2-FxNLMS and \mbox{R-CC-FxCNLMS} algorithms respectively. The kernel width is $\sigma=1$ in these algorithms.  Fig. \ref{Fig13} evaluates the performance of the \mbox{R-FxNLMS}, \mbox{SSS-FxNLMS}, proposed \mbox{R-SSS-FxNLMS} algorithms. The performance of the \mbox{SSS-FxNLMS} algorithm decreases in impulsive noise environment, while the \mbox{R-SSS-FxNLMS} algorithm has a more stable performance due to its robust strategy. Moreover, similar the \mbox{SSS-FxNLMS} algorithm, the \mbox{R-SSS-FxNLMS} algorithm can also achieve fast convergence rate of large step-size and low steady-state residual error of small step-size. Alos the step-size switching process of the \mbox{R-SSS-FxNLMS} algorithm is shown in Fig. \ref{Fig14}. \mbox{Fig. \ref{Fig15}} depicts the ANR curves of different algorithms. Obviously, all of the four algorithms are robust and the proposed R-SSS-FxNLMS algorithm shows clear superiority over the other robust algorithms.

In order to further explore the universality and performance differences of various robust functions in suppressing impulsive noise interference, we incorporate the robustness strategy of EHCF\cite{kumar2024robust, 9470980} into the update \eqref{eq027} of \mbox{R-SSS-FxNLMS}. Hence, the robust function $\varphi [e_{m}]$ and its scaling factor $g[e_m]$ are formulated as, respectively,
\begin{equation}
	\varphi [e_{m}]=1-\text{E}[\beta \text{exp}[-|\text{cosh}(\eta e_{m})|^{\theta}]],
	\label{eq038}
\end{equation}
and
\begin{equation}
	\begin{aligned}
	g[e_{m}]=&\text{exp}\left\{-|\text{cosh}|[\eta e_{m}]^{\theta}\right\}\text{sinh}[\eta e_{m}]\cosh[\eta e_{m}]^{\theta-1}\\
	&\text{sign}\left\{\text{cosh}[\eta e_{m}]\right\}/e_{m}.
	\label{eq039}
	\end{aligned}
\end{equation}
                   
\begin{figure}[htbp]
	\centering
	{
		\begin{minipage}[a]{0.4\textwidth}
			\includegraphics[width=1\textwidth]{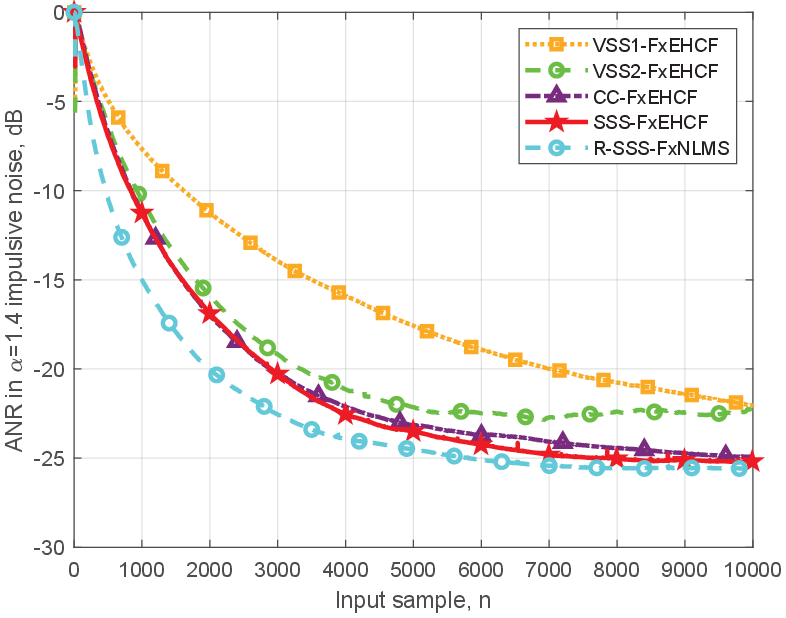} 
		\end{minipage}
	}
	\caption{ANR curves of different algorithms in $\alpha=1.4$ impulsive noise.}
	\label{Fig16}
\end{figure}

Similarly, we also incorporate the robustness strategy of EHCF into the VSS1-FxNLMS, VSS2-FxNLMS, \mbox{CC-FxCNLMS},  and SSS-FxNLMS algorithms, respectively yielding the  VSS1-FxEHCF, VSS2-FxEHCF, \mbox{CC-FxEHCF}, and \mbox{SSS-FxEHCF} algorithms. Fig. \ref{Fig16} illustrates the ANR curves of these EHCF-based algorithms. Evidently, four EHCF-based algorithms demonstrate robustness against the $\alpha$-stable noise, and the proposed SSS-FxEHCF algorithm exhibits superior performance compared to other EHCF-based algorithms. In comparison, the R-SSS-FxNLMS algorithm using the MCC strategy could achieve a faster convergence than the SSS-FxEHCF algorithm.

\begin{figure}[htbp]
	\centering
	{
		\begin{minipage}[a]{0.4\textwidth}
			\includegraphics[width=1\textwidth]{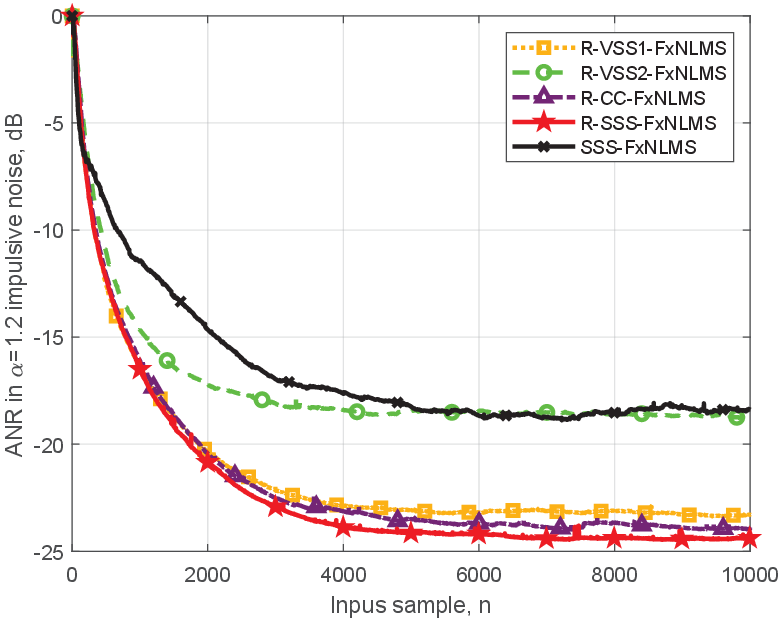} 
		\end{minipage}
	}
	\caption{ANR curves of different algorithms in $\alpha=1.2$ impulsive noise.}
	\label{Fig17}
\end{figure}

To further verify the robustness of the proposed algorithm against impulsive noise, additional simulations were conducted with $\alpha=1.2$ and $\gamma=0.3$. As shown in Fig. \ref{Fig17}, all algorithms maintain stable performance under these impulsive noise settings, and the R-SSS-FxNLMS algorithm continues to achieve excellent results.

\subsection{Experiment 3: factory noise}

\begin{figure}[htbp]
	\centering
	{
		\begin{minipage}[a]{0.4\textwidth}
			\includegraphics[width=1\textwidth]{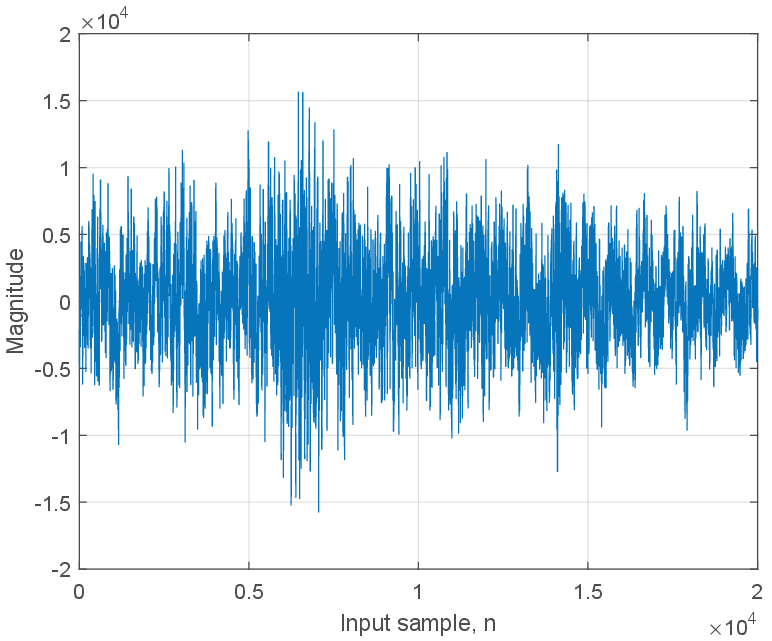} 
		\end{minipage}
	}
	\caption{Factory floor noise.}
	\label{Fig18}
\end{figure}

To prove the practicality of proposed SSS-FxNLMS algorithm in actual noise environments, experiments are conducted using factory floor noise. It is sourced from the public data of the Signal Processing Information Base\footnote{http://spib.linse.ufsc.br/noise.html}, and 
the \mbox{amplitude-frequency} characteristics of the noise are shown in Fig. \ref{Fig18}. Specially, the rule for setting the step-sizes is that $\mu_{k}=0.6\mu_{k-1}$. All the experiments were carried out only once to obtain the results.

\begin{figure}[htbp]
	\centering
	{
		\begin{minipage}[a]{0.4\textwidth}
			\includegraphics[width=1\textwidth]{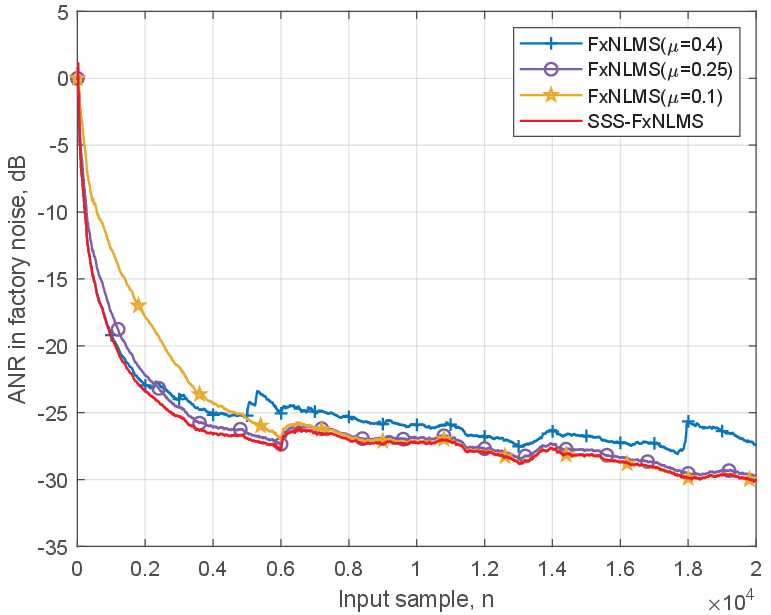} 
		\end{minipage}
	}
	\caption{ANR curves of the SSS-FxNLMS algorithm and standard FxNLMS algorithm with three fixed step-sizes in factory floor noise.}
	\label{Fig19}
\end{figure}

\begin{figure}[htbp]
	\centering
	{
		\begin{minipage}[a]{0.4\textwidth}
			\includegraphics[width=1\textwidth]{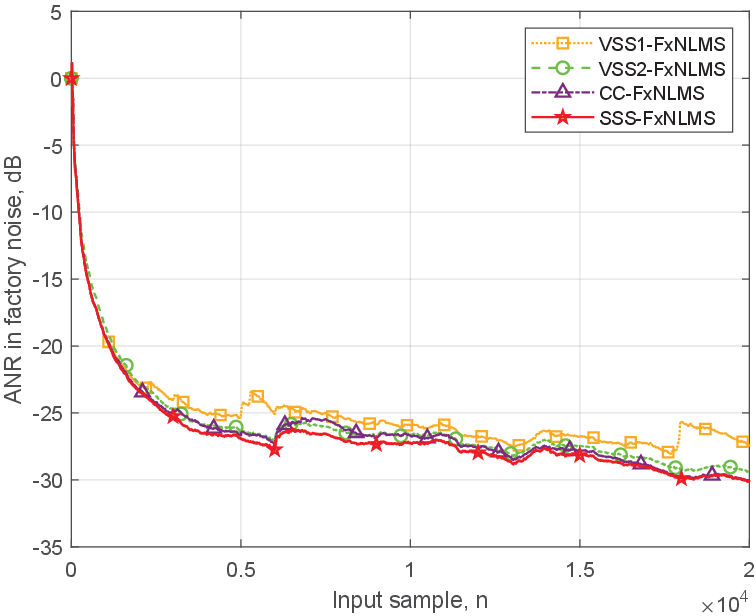} 
		\end{minipage}
	}
	\caption{ANR curves of different algorithms in factory floor noise.}
	\label{Fig20}
\end{figure}

We verify the performance of proposed SSS-FxNLMS algorithm in the context of the factory noise in Figs. \ref{Fig19} and \ref{Fig20}. It is obvious that SSS-FxNLMS algorithm has advantages of both large and small step-sizes, and compared with the other three comparative algorithms, it has superior performance.

\subsection{Experiment 4: hammering noise}

\begin{figure}[htbp]
	\centering
	{
		\begin{minipage}[a]{0.4\textwidth}
			\includegraphics[width=1\textwidth]{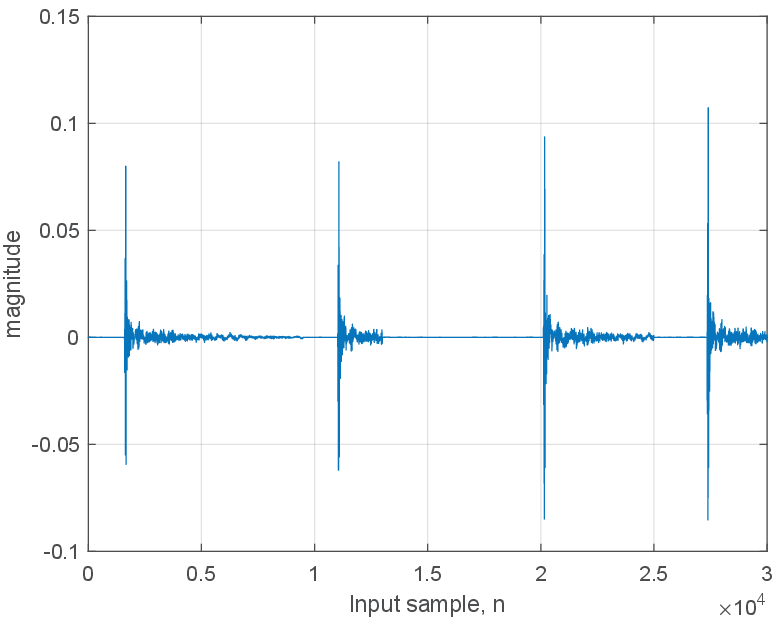} 
		\end{minipage}
	}
	\caption{Hammering noise.}
	\label{Fig21}
\end{figure}
\begin{figure}[t!]
	\centering
	{
		\begin{minipage}[a]{0.4\textwidth}
			\includegraphics[width=1\textwidth]{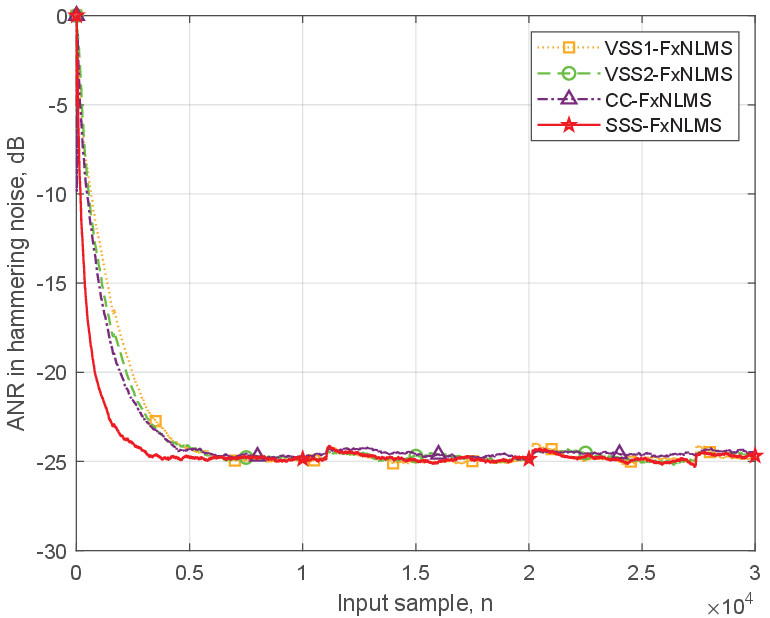} 
		\end{minipage}
	}
	\caption{ANR curves of different algorithms in hammering floor noise.}
	\label{Fig22}
\end{figure}

Hammering noise\footnote{https://freesound.org/people/SebastianVDB/sounds/653162/} is a special case of impulsive noises, which exhibits a repetitive impulsive characteristic in the time-domain but has a relatively low amplitude as compared to the $\alpha$-stable noise, as shown in Fig. \ref{Fig21}. Considering the background noise in practice, we refer to the method in \cite{sun2015convergence} and add a white Gaussian noise with variance of $0.001$ to the pure hammering noise signal. It has been proved in \cite{sun2015convergence} that the FxLMS algorithm is applicable to the processing of such noise. Moreover, the FxNLMS algorithm owns suppression capability against impulsive noises due to its normalization mechanism for the reference input impulsive. Therefore, for the case of using the hammering noise as the reference input, in Fig. \ref{Fig22} we give the ANR results of those algorithms in Fig. \ref{Fig5}. As one can see, several algorithms have shown favorable noise reduction, while the proposed SSS-FxNLMS algorithm has faster convergence rate.

\section{Conclusion}
In this paper, to resolve the conflict between the convergence rate and steady-state residual error of the FxNLMS algorithm, we extended the switched step-size idea to ANC. We analyzed and established the MSD recursion of the FxNLMS algorithm and developed the SSS-FxNLMS algorithm. It can achive both the fast convergence rate of a large step-size and the small steady-state residual error of a small step-size. The advantage of this algorithm is that it only needs the MSD trend instead of the exact value, which facilitates its implementation. Furthermore, to improve the robustness of the \mbox{SSS-FxNLMS} algorithm, we proposed its robust variants by incorporating the specified robust functions. The superiority of the proposed algorithms have been verified in different noise scenarios.

\appendices
\section{\break Derivation of $\left(14\right)$}
\renewcommand{\theequation}{A.\arabic{equation}}

\setcounter{equation}{0}

Putting $\left(\ref{eq008}\right)$ into the second term of $\left(\ref{eq013}\right)$, it yields
\begin{equation}
	\begin{aligned}
		\mu\text{E}\left\{\frac{\textbf{x}_{f,m}e_{m}}{||\textbf{x}_{f,m}||_{2}^{2}}\widetilde{\textbf{w}}^{\text{T}}_{m}\right\}=&\mu\text{E}\left\{\frac{\textbf{x}_{f,m}\textbf{x}_{f,m}^{\text{T}}\widetilde{\textbf{w}}_{m}}{||\textbf{x}_{f,m}||_{2}^{2}}\widetilde{\textbf{w}}^{\text{T}}_{m}\right\}\\
		&+\mu\text{E}\left\{\frac{\textbf{x}_{f,m}v_{m}}{||\textbf{x}_{f,m}||_{2}^{2}}\widetilde{\textbf{w}}^{\text{T}}_{m}\right\}.
		\label{eqA1}
	\end{aligned}
\end{equation}
Considering the assumption 1, the second term of \eqref{eqA1} can be simplified as
\begin{equation}
	\begin{aligned}
		\mu\text{E}\left\{\frac{\textbf{x}_{f,m}v_{m}}{||\textbf{x}_{f,m}||_{2}^{2}}\widetilde{\textbf{w}}^{\text{T}}_{m}\right\}=0.
		\label{eqA2}
	\end{aligned}
\end{equation}
Substituting \eqref{eqA2} into \eqref{eqA1} and using assumption 2, we can obtain
\begin{equation}
	\begin{aligned}
		&\mu\text{E}\left\{\frac{\textbf{x}_{f,m}e_{m}}{||\textbf{x}_{f,m}||_{2}^{2}}\widetilde{\textbf{w}}^{\text{T}}_{m}\right\}=\mu\frac{\textbf{R}_{f,m}\textbf{P}_{m}}{\text{E}\left\{||\textbf{x}_{f,m}||_{2}^{2}\right\}}
		\label{eqA3}
	\end{aligned}
\end{equation}
where $\textbf{R}_{f,m}\triangleq\text{E}\left\{\textbf{x}_{f,m}\textbf{x}_{f,m}^{\text{T}}\right\}$.

Taking advantage of the same method to \eqref{eqA3}, the third term of $\left(\ref{eq013}\right)$ can be simplified as
\begin{equation}
	\begin{aligned}
		\mu\text{E}\left\{\widetilde{\textbf{w}}_{m}\frac{\textbf{x}_{f,m}^{\text{T}}e_{m}}{||\textbf{x}_{f,m}||_{2}^{2}}\right\}=\mu\frac{\textbf{P}_{m}\textbf{R}_{f,m}}{\text{E}\left\{||\textbf{x}_{f,m}||_{2}^{2}\right\}}.
		\label{eqA4}
	\end{aligned}
\end{equation}

By inserting $\left(\ref{eq008}\right)$ into the fourth term of $\left(\ref{eq013}\right)$ which yields
\begin{equation}
	\begin{aligned}
		\mu^{2}\text{E}\left\{e^{2}_{m}\frac{\textbf{x}_{f,m}\textbf{x}_{f,m}^{\text{T}}}{||\textbf{x}_{f,m}||_{2}^{4}}\right\}=&\mu^{2}\text{E}\left\{\frac{\textbf{x}_{f,m}\textbf{x}_{f,m}^{\text{T}}\widetilde{\textbf{w}}_{m}\widetilde{\textbf{w}}^{\text{T}}_{m}\textbf{x}_{f,m}\textbf{x}_{f,m}^{\text{T}}}{||\textbf{x}_{f,m}||_{2}^{4}}\right\}\\
		&+\mu^{2}\text{E}\left\{\frac{\textbf{x}_{f,m}\textbf{x}_{f,m}^{\text{T}}\widetilde{\textbf{w}}_{m}v_{m}\textbf{x}_{f,m}^{\text{T}}}{||\textbf{x}_{f,m}||_{4}^{2}}\right\}\\
		&+\mu^{2}\text{E}\left\{\frac{\textbf{x}_{f,m}v_{m}\widetilde{\textbf{w}}^{\text{T}}_{m}\textbf{x}_{f,m}\textbf{x}_{f,m}^{\text{T}}}{||\textbf{x}_{f,m}||_{4}^{2}}\right\}\\
		&+\mu^{2}\sigma_{v}^{2}\text{E}\left\{\frac{\textbf{x}_{f,m}\textbf{x}_{f,m}^{\text{T}}}{||\textbf{x}_{f,m}||_{2}^{4}}\right\}.
		\label{eqA5}
	\end{aligned}
\end{equation}
Based on the Gaussian Matrix Decomposition Theorem\cite{wu1997adaptive,yin2010stochastic}, the following equivalent substitution is available:
\begin{equation}
	\begin{aligned}
		\text{E}\left\{\textbf{x}_{f,m}\textbf{x}_{f,m}^{\text{T}}\widetilde{\textbf{w}}_{m}\widetilde{\textbf{w}}^{\text{T}}_{m}\textbf{x}_{f,m}\textbf{x}_{f,m}^{\text{T}}\right\}=&2\textbf{R}_{f,m}\textbf{P}_{m}\textbf{R}_{f,m}\\
		&+\textbf{R}_{f,m}\text{Tr}[\textbf{R}_{f,m}\textbf{P}_{m}],
		\label{eqA6}
	\end{aligned}
\end{equation}
where $\text{Tr}[\cdot]$ is denoted as the trace of a matrix.

According to assumption 1 and 2, the second and third terms of \eqref{eqA5} can be simplified to respectively:
\begin{equation}
	\mu^{2}\text{E}\left\{\frac{\textbf{x}_{f,m}\textbf{x}_{f,m}^{\text{T}}\widetilde{\textbf{w}}_{m}v_{m}\textbf{x}_{f,m}^{\text{T}}}{||\textbf{x}_{f,m}||_{4}^{2}}\right\}=0
	\label{eqA7}
\end{equation}
and
\begin{equation}
	\mu^{2}\text{E}\left\{\frac{\textbf{x}_{f,m}v_{m}\widetilde{\textbf{w}}^{\text{T}}_{m}\textbf{x}_{f,m}\textbf{x}_{f,m}^{\text{T}}}{||\textbf{x}_{f,m}||_{4}^{2}}\right\}=0.
	\label{eqA8}
\end{equation}

Let us substitute \eqref{eqA6}-\eqref{eqA8} into \eqref{eqA5} and then it can be simplified as
\begin{equation}
	\begin{aligned}
		&\mu^{2}\text{E}\left\{e^{2}_{m}\frac{\textbf{x}_{f,m}\textbf{x}_{f,m}^{\text{T}}}{||\textbf{x}_{f,m}||_{2}^{4}}\right\}=\mu^{2}\frac{\sigma_{v}^{2}\textbf{R}_{f,m}}{\text{E}\left\{||\textbf{x}_{f,m}||_{2}^{4}\right\}}\\
		&+\mu^{2}\frac{2\textbf{R}_{f,m}\textbf{P}_{m}\textbf{R}_{f,m}+\textbf{R}_{f,m}\text{Tr}\left\{\textbf{R}_{f,m}\textbf{P}_{m}\right\}}{\text{E}\left\{||\textbf{x}_{f,m}||_{2}^{4}\right\}}.
		\label{eqA9}
	\end{aligned}
\end{equation}

Based on \eqref{eqA3}, \eqref{eqA4} and \eqref{eqA9}, wee can rewrite \eqref{eq013} as
\begin{equation}
	\begin{aligned}
		\textbf{P}_{m+1}=&\textbf{P}_{m}-2\mu\frac{\textbf{R}_{f,m}\textbf{P}_{m}}{\text{E}\left\{||\textbf{x}_{f,m}||_{2}^{2}\right\}}+2\mu^{2}\frac{\textbf{R}_{f,m}\textbf{P}_{m}\textbf{R}_{f,m}}{\text{E}\left\{||\textbf{x}_{f,m}||_{2}^{4}\right\}}\\
		&+\mu^{2}\frac{\textbf{R}_{f,m}\text{Tr}[\textbf{R}_{f,m}\textbf{P}_{m}]}{\text{E}\left\{||\textbf{x}_{f,m}||_{2}^{4}\right\}}+\mu^{2}\frac{\sigma_{v}^{2}\textbf{R}_{f,m}}{\text{E}\left\{||\textbf{x}_{f,m}||_{2}^{4}\right\}}.
		\label{eqA10}
	\end{aligned}
\end{equation}

The fifth term of \eqref{eqA10} involves the variance of the measurement noise $v_{m}$, but it usually cannot be directly obtained. With the adaptation of the filter weights, $\widetilde{\textbf{w}}_{m}$ will approach $0$ gradually so that $e_{m}$ is gradually close $v_{m}$. So, we consider approximating $\sigma_{v}^{2}$ with $\sigma^{2}_{e,m}\triangleq\text{E}\left\{e^{2}_{m}\right\}$ and then \eqref{eqA10} is changed to

\begin{equation}
	\begin{aligned}
		\textbf{P}_{m+1}=&\textbf{P}_{m}-2\mu\frac{\textbf{R}_{f,m}\textbf{P}_{m}}{\text{E}\left\{||\textbf{x}_{f,m}||_{2}^{2}\right\}}+2\mu^{2}\frac{\textbf{R}_{f,m}\textbf{P}_{m}\textbf{R}_{f,m}}{\text{E}\left\{||\textbf{x}_{f,m}||_{2}^{4}\right\}}\\
		&+\mu^{2}\frac{\textbf{R}_{f,m}\text{Tr}[\textbf{R}_{f,m}\textbf{P}_{m}]}{\text{E}\left\{||\textbf{x}_{f,m}||_{2}^{4}\right\}}+\mu^{2}\frac{\sigma_{e}^{2}\textbf{R}_{f,m}}{\text{E}\left\{||\textbf{x}_{f,m}||_{2}^{4}\right\}}.
		\label{eqA11}
	\end{aligned}
\end{equation}
where the power $\sigma^{2}_{e,m}$ is calculated in a recursive estimate way,
\begin{equation}
	\begin{aligned}
		\sigma^{2}_{e,m}=\lambda\sigma^{2}_{e,m}+(1-\lambda){e}^{2}_{e,m},
	\end{aligned}
	\label{eqA12}
\end{equation}
with $0<\lambda<1$ being the forgetting factor.
Moreover, the expectations on $\textbf{x}_{f}$ in \eqref{eqA11}, we choose to instead them with their instantaneous values \cite{guo2022normalized}. As a result, we have completed the derivation of \eqref{eqA11}.

\end{document}